\documentclass[pra,twocolumn,nofootinbib,floatfix,10pt]{revtex4-2}

\usepackage{amsmath}
\usepackage{amssymb}
\usepackage{wasysym}
\usepackage{graphicx}
\usepackage{color,soul}
\usepackage{physics}
\usepackage{siunitx}
\usepackage{dsfont}
\usepackage{float}
\usepackage{xcolor}
\usepackage[english]{babel}
\usepackage{blindtext}
\usepackage[english,nomargin,inline,marginclue,draft]{fixme}
\pdfpageheight\paperheight
\pdfpagewidth\paperwidth

\usepackage[colorlinks,linkcolor=blue,anchorcolor=blue,citecolor=blue,urlcolor=blue]{hyperref}

\fxusetheme{colorsig}
\FXRegisterAuthor{cg}{acg}{CG}  
\FXRegisterAuthor{th}{ath}{\color{blue}TH}  
\FXRegisterAuthor{ib}{aib}{\color{red}IB} 
\FXRegisterAuthor{sh}{ash}{\color{cyan}SH} 
\FXRegisterAuthor{db}{adb}{\color{green}DB} % now I can ..
\FXRegisterAuthor{ps}{aps}{PS}
\makeatletter
\renewcommand*\FXLayoutInline[3]{%
  {\@fxuseface{inline}\ignorespaces{\color{fx#1}[#3: #2]}}}
\makeatother

\long\def\symbolfootnote[#1]#2{\begingroup%
\def\thefootnote{\fnsymbol{footnote}}\footnotetext[#1]{#2}\endgroup}

\def\nobreakbefore{%
  \relax\ifvmode\else
    \ifhmode
      \ifdim\lastskip > 0pt\relax
        \unskip\nobreakspace
      \else % added to put a ~if no space was typed. (Unclear why it sometimes worked before )
        \nobreakspace
      \fi
    \fi
  \fi
}
\let\oldcite\cite
\renewcommand\cite{\nobreakbefore\oldcite}

\begin{document}
\title{Enhanced multi-parameter metrology in dissipative Rydberg atom time crystals}

\author{Bang Liu$^{1,2,\textcolor{blue}{\star}}$}
\author{Jun-Rong Chen$^{3,\textcolor{blue}{\star}}$}
\author{Yu Ma$^{1,2,\textcolor{blue}{\star}}$}
\author{Qi-Feng Wang$^{1,2,\textcolor{blue}{\star}}$}
\author{Tian-Yu Han$^{1,2}$}
\author{Hao Tian$^{3}$}
\author{Yu-Hua Qian$^{4}$}
\author{Guang-Can Guo$^{1,2}$}
\author{Li-Hua Zhang$^{1,2,\textcolor{blue}{\#}}$}
\author{Bin-Bin Wei$^{5,\textcolor{blue}{\ddagger}}$}
\author{Abolfazl Bayat$^{6,\textcolor{blue}{\S}}$}
\author{Dong-Sheng Ding$^{1,2,\textcolor{blue}{\dagger}}$}
\author{Bao-Sen Shi$^{1,2}$}

\affiliation{$^1$Key Laboratory of Quantum Information, University of Science and Technology of China; Hefei, Anhui 230026, China.}
\affiliation{$^2$Anhui Province Key Laboratory of Quantum Network, University of Science and Technology of China, Hefei 230026, China.}
\affiliation{$^3$School of physics, Harbin Institute of Technology, Harbin, Heilongjiang 150001, China.}
\affiliation{$^4$Institute of Big Data Science and Industry, Shanxi University, Taiyuan, China}
\affiliation{$^5$Institute of system engineering, Tianjin 300161, China.}
\affiliation{$^6$Institute of Fundamental and Frontier Sciences, University of Electronic Science and Technology of China, Chengdu 611731, China}

\date{\today}

\symbolfootnote[1]{B.L., J.R.C., Y.M. and Q.F.W. contribute equally to this work.}
\symbolfootnote[2]{zlhphys@ustc.edu.cn}
\symbolfootnote[3]{weibb.2009@tsinghua.org.cn}
\symbolfootnote[4]{abolfazl.bayat@uestc.edu.cn}
\symbolfootnote[2]{dds@ustc.edu.cn}

\maketitle

\textbf{The pursuit of unprecedented sensitivity in quantum enhanced metrology has spurred interest in non-equilibrium quantum phases of matter and their symmetry breaking. In particular, criticality-enhanced metrology through time-translation symmetry breaking in many-body systems, a distinct paradigm compared to spatial symmetry breaking, is a field still in its infancy. Here, we have investigated the enhanced sensing at the boundary of a continuous time-crystal (CTC) phase in a driven Rydberg atomic gas. By mapping the full phase diagram, we identify the parameter-dependent phase boundary where the time-translation symmetry is broken. This allows us to use a single setup for measuring multiple parameters, in particular frequency and amplitude of a microwave field. By increasing the microwave field amplitude, we first observe a phase transition from a thermal phase to a CTC phase, followed by a second transition into a distinct CTC state, characterized by a different oscillation frequency. Furthermore, we reveal the precise relationship between the CTC phase boundary and the scanning rate, displaying enhanced precision beyond the Standard Quantum Limit. This work not only provides a promising paradigm rooted in the critical properties of time crystals, but also advances a method for multi-parameter sensing in non-equilibrium quantum phases. }

Advancing quantum metrology not only pushes the frontiers of fundamental physics but also catalyzes the development of next-generation measurement technologies across diverse fields \cite{giovannetti2011advances,montenegro2025quantum}. The critical points associated with symmetry breaking are known to amplify a system's response to external perturbations, providing a powerful paradigm in quantum enhanced metrology \cite{gammelmark2011phase, macieszczak2016dynamical, PhysRevA.96.013817,raghunandan2018high,PhysRevLett.124.120504,PhysRevLett.126.200501,PhysRevLett.126.010502,PhysRevLett.126.200501,Theodoros2021Criticality,garbe2021critical,liu2021experimental,PhysRevA.78.042105,sarkar2025exponentially,sarkar2024critical}. The sensing technique operates by applying a minute perturbation to a many-body system at its critical point, where the system's divergent susceptibility acts as an amplifier, converting the microscopic input into a macroscopic, detectable signal with enhanced precision \cite{yang2022variational,montenegro2025quantum}. However, current experiments have predominantly focused on sensing a single parameter~\cite{liu2021experimental,ding2022enhanced,moon2024discrete}. This presents a significant gap, as real-world sensing scenarios, from biomedical diagnostics to environmental monitoring and fundamental physics experiments, are inherently multi-parametric, involving the simultaneous and often correlated measurement of multiple physical quantities \cite{gessner2018sensitivity,yang2019optimal,albarelli2022probe,valeri2023experimental,kaubruegger2023optimal}.

\begin{figure*}
    \centering
    \includegraphics[width=2.08\columnwidth]{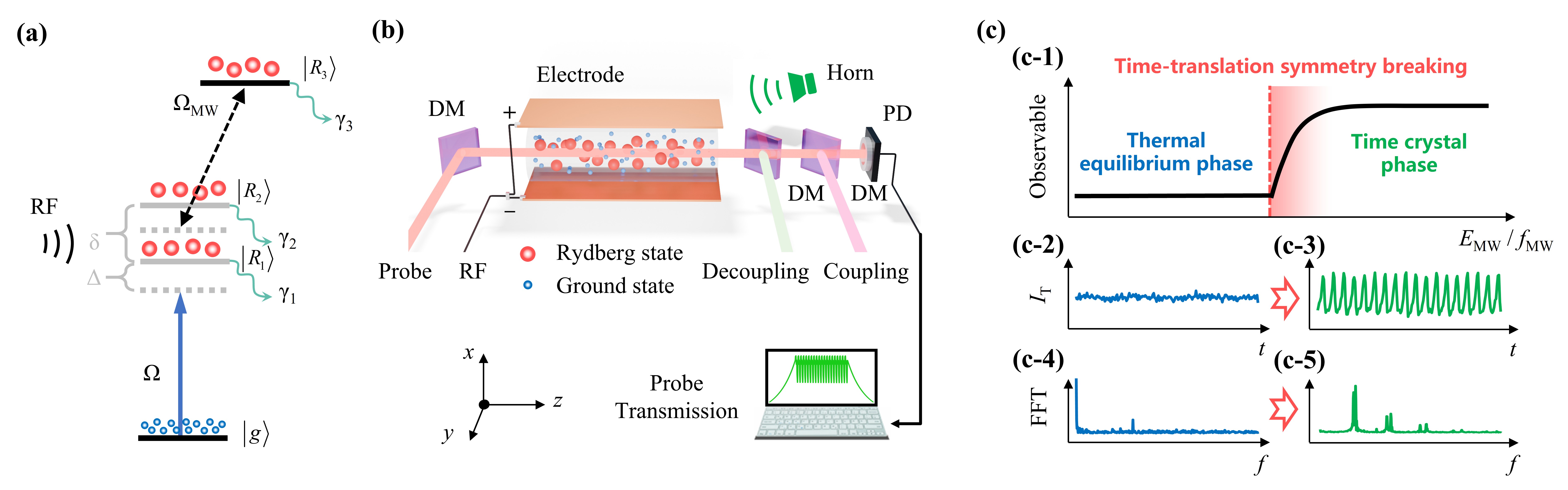}
    \caption{\textbf{Experimental diagram and the criticality enhanced metrology model.} (a) Energy level diagram of model. The level structure consists of the atomic ground state $\ket{g}$and three Rydberg states $\ket{R_1}$, $\ket{R_2}$ and $\ket{R_3}$. (b) Schematic diagram of the experimental setup. The experiment employs a three-photon Rydberg excitation scheme. The probe field propagates in the opposite direction to the dressing field and the coupling field, passing through the atomic vapor cell and finally being received by a photo-detector. Electrodes are used to radiate RF electric field, while antennas are used to generate microwave electric field. (c) The criticality enhanced metrology model. As the microwave amplitude (or its frequency) is varied, the system undergoes a phase transition from a thermal equilibrium phase to a time crystal phase. Near the critical point of the phase transition, the system exhibits higher sensitivity to external perturbations, which can be exploited for enhanced metrology. (c2) and (c4) represent the measured probe transmission in time and frequency domain in the thermal equilibrium phase. (c3) and (c5) correspond to the cases in the time crystal phase.}

    \label{fig1}
\end{figure*}

Time crystals have emerged as a prominent class of nonequilibrium phases of matter in which time-translational symmetry is spontaneously broken. They emerge in two distinct categories: (i) discrete time crystals in periodically driven Floquet systems where the system's response is locked to an integer multiple of the driving period~\cite{sacha2017time,sacha2020time,zaletel2023colloquium}; and (ii) continuous time crystals in dissipative systems where long-lasting oscillations appear with no sign of equilibration~\cite{iemini2018boundary}. 
Both discrete~\cite{zhang2017observation,choi2017observation,gong2018discrete,else2016floquet,PhysRevA.91.033617,gambetta2019discrete,huang2018clean,rovny2018observation,Frey2022,Mi2021,randall2021many,Xiang2024} and continuous~\cite{autti2018observation,smits2018observation,kongkhambut2022observation,iemini2018boundary,kessler2021observation,CarraroHaddad2024Solid-state,LiYaohua2024,Huang2025,Greilich2024,Chen2023,Zhao2025Space-time} time crystals have been experimentally observed in various physical systems. Recently, a new application for time crystals have been proposed to exploit them as quantum sensors with potential advantage over classical probes~\cite{cabot2024continuous,o2025quantum,iemini2018boundary,yousefjani2025discrete,Iemini2024floquet}. However, experimental realization of such probes is only limited to discrete time crystals in solid state system with no quantum enhancement~\cite{moon2024discrete}. From an experimental perspective, this leads to three central questions: Can CTCs function as practical quantum sensors? Can a single CTC probe be configured to measure multiple parameters? And, crucially, can such a system demonstrate experimentally observed sensitivity beyond the Standard Quantum Limit (SQL)?

Driven-dissipative ensembles of Rydberg atoms are excellent candidates to study quantum many-body physics. In this context, the long-range Rydberg interaction between atoms allow us to investigate criticality and phase transitions~\cite{lee2012collective,carr2013nonequilibrium,helmrich2020signatures,ding2019Phase,wadenpfuhl2023emergence,ding2023ergodicity,zhang2024early,zhang2025exceptional}, dissipative time crystals~\cite{wu2024observation,liu2025bifurcation,jiao2025observation}, higher-order and fractional discrete time crystals~\cite{liu2024higher} as well as discrete time quasicrystals~\cite{zhu2025observation}. In the context of quantum sensing, thanks to their large dipole moments~\cite{gallagher2005Rydberg,firstenberg2016nonlinear}, the Rydberg atoms have also been used as a probe for measuring microwave fields~\cite{sedlacek2012microwave,jing2020atomic,liuHighly,zhang2022rydberg,zhang2024ultra,liu2023electric,zhang2024rydberg,liu2025cavity,wang2025measurement} enabling criticality-enhanced electric field sensing~\cite{wade2018terahertz,ding2022enhanced,wang2025high}. Merging these two distinct applications into a single setup, namely experimentally realizing a Rydberg-based quantum sensor in a dissipative time-crystal phase, is highly desirable. Such a sensor would be particularly powerful if capable of detecting multiple parameters with precision beyond the SQL.

In this work, we have demonstrated an experiment of quantum enhanced metrology near the critical point of time-translation symmetry breaking in driven-dissipative ensembles of Rydberg atoms. Our setup can be used for detecting both frequency and amplitude of a microwave field. The enhanced sensitivity results from the system's operation at a non-equilibrium critical point, where the interplay between long-range interactions and dissipation leads to a divergent susceptibility. Through measurements of the probe transmission versus microwave field amplitude, we observe a criticality-enhanced response featuring two distinct criticality regimes. These criticalities correspond to the phase transition from a thermal equilibrium phase to a CTC phase and the subsequent transition between distinct CTC phases, each characterized by unique scaling behavior and critical slowing down. Our work demonstrates a dramatic enhancement in precision at criticality in multi-parameter measurement of microwave frequency and amplitude, supporting an improvement of over two orders of magnitude compared to the thermal equilibrium phase and achieving a precision well beyond the SQL. Our work establishes a new paradigm for quantum sensing that leverages the critical properties of non-equilibrium phase transitions and opens a pathway for exploiting temporal order in metrological applications.

\begin{figure*}
    \centering
    \includegraphics[width=2.08\columnwidth]{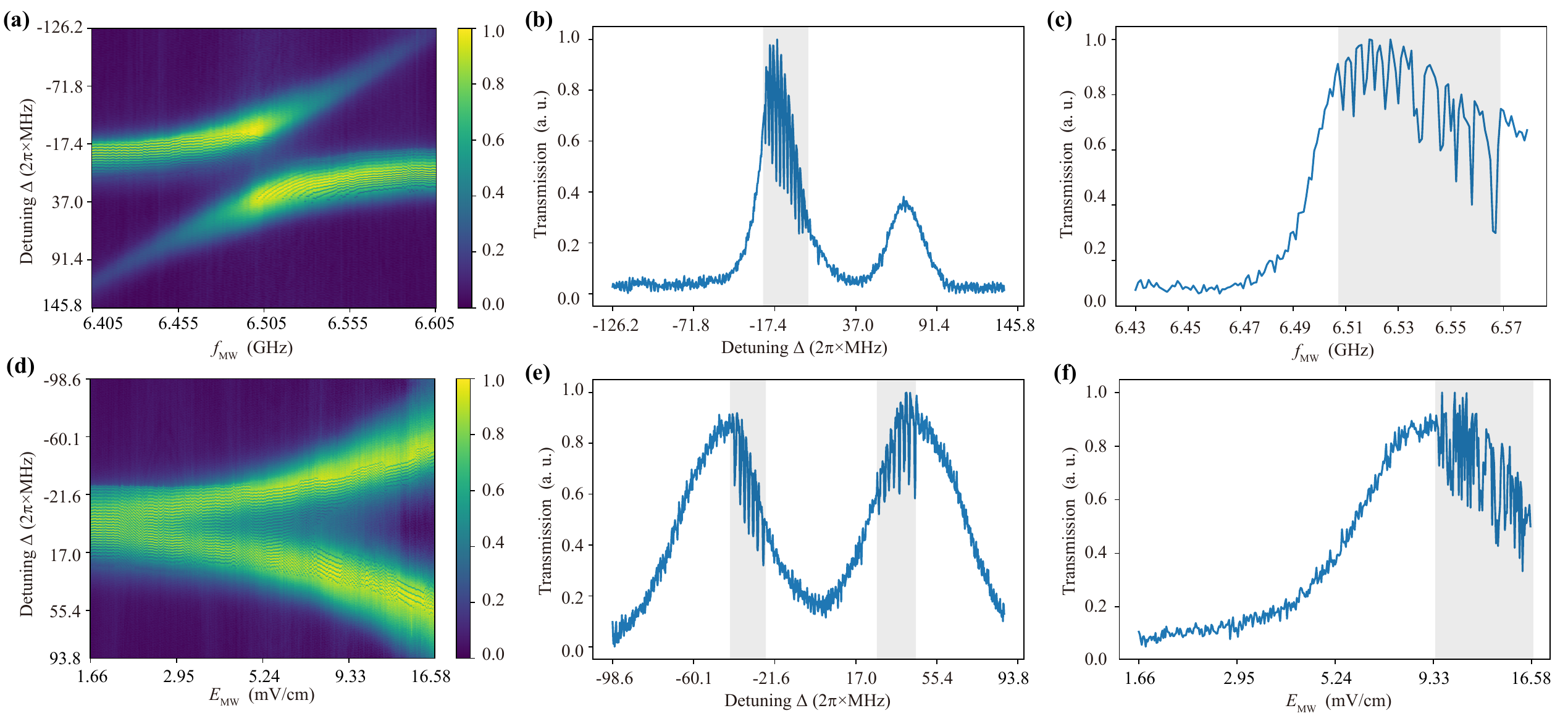} 
    \caption{\textbf{Measured phase diagrams.} (a) Transmission spectrum obtained by sweeping the microwave frequency $f_{\rm{MW}}$ from 6.405 to 6.605 GHz. The coupling of the microwave field to the Rydberg states induces a pronounced splitting of the transmission resonance into two distinct peaks, accompanied by coherent spectral oscillations that signal the emergence of the time crystal phase. (b) Spectrum recorded at a fixed microwave frequency $f_{\rm{MW}} = 6.455$ GHz and field amplitude $E_{\rm{MW}} = 9.33~ \rm{mV/cm}$. (c) Normalised probe transmission as a function of $f_{\rm{MW}}$ under conditions of fixed detuning, $\Delta=2\pi \times 23.4~\rm{MHz}$, and fixed microwave amplitude, $E_{\rm{MW}} = 9.33~ \rm{mV/cm}$. (d) Evolution of the transmission spectrum with increasing microwave field amplitude $E_{\rm{MW}}$ (from 1.66 to 16.58 mV/cm) under resonant drive ($f_{\rm{MW}} = 6.505$ GHz). The peak splitting widens progressively with $E_{\rm{MW}}$, and the spectral region associated with the time crystal phase also bifurcates. (e) Spectrum corresponding to $E_{\rm{MW}} = 9.33~\rm{mV/cm}$. (f) Phase transition observed by varying $E_{\rm{MW}}$  while maintaining a constant detuning of $\Delta=-2\pi\times37.9 $ MHz, demonstrating a crossover from a thermal to a time crystal phase. The colour bar represents the normalised probe transmission intensity. Grey-shaded regions in all panels denote the parameter space where the time crystal phase is stabilised.}
    \label{fig2}
\end{figure*}

\subsection*{Physical model}
    To investigate the criticality enhanced metrology, we consider a many-body atomic system comprising $N$ atoms with a ground state $\ket{g}$, and two nearly degenerate Rydberg states $\ket{R_1}$ and $\ket{R_2}$ (with an energy interval $\delta$), as depicted in Fig.~\ref{fig1}(a). The coupling between ground state $\ket{g}$ and the Rydberg state $\ket{R_1}$ (or $\ket{R_2}$) is characterized by the Rabi frequency $\Omega_1$($\Omega_2$) and detuning $\Delta$($\Delta+\delta$). The microwave field couples the two Rydberg states $\ket{R_1}$ and $\ket{R_2}$ to the other Rydberg states $\ket{R_3}$ with a Rabi frequency $\Omega_{\text{MW}}$ and detuning $\Delta_{\text{MW}}$. The Hamiltonian is described by double Rydberg state model with microwave driving~\cite{wu2024observation,liu2024higher}: 
\begin{equation}
    \begin{aligned}
        \hat{H} & =\frac{1}{2}\sum_{i}\left(\Omega_{1}\sigma_{i}^{gR_1}+\Omega_{2}\sigma_{i}^{gR_2}+\Omega_{\text{MW}}\sigma_{i}^{R_2R_3}+\Omega_{\text{MW}}\sigma_{i}^{R_1R_3}\right.\\ &\left.+\rm{H.c.}\right)-\sum_{i}\left(\Delta n_{i}^{R_1}+\Delta_1 n_{i}^{R_2}+\Delta_2 n_{i}^{R_3}\right) \\ &+\sum_{i\neq j}V_{ij}\bigg[n_{i}^{R_1}n_{j}^{R_2}+n_{i}^{R_2}n_{j}^{R_3}+n_{i}^{R_1}n_{j}^{R_3}\\ &+\frac{1}{2}(n_{i}^{R_1}n_{j}^{R_1}+n_{i}^{R_2}n_{j}^{R_2}+n_{i}^{R_3}n_{j}^{R_3})\bigg]
    \end{aligned}\label{Hamiltonian}
    \end{equation}
where $\sigma_{i}^{gr}$ ($r \in {R_1,R_2}$) denotes the transition operator between the ground state  $\left| g \right\rangle$ and the Rydberg state $\left|  r \right\rangle$ for the $i$-th atom, $n_{i}^{R_1,R_2,R_3}$ represent the population operators for the respective Rydberg levels $\left|  R_1 \right\rangle$, $\left|  R_2 \right\rangle$ and $\left|  R_3 \right\rangle$, $\Delta_1=\Delta +\delta$, $\Delta_2=\Delta + \Delta_\text{MW} +\delta$, and $V_{ij}$ describes the Rydberg-Rydberg interaction between atoms $i$ and $j$.

The criticality-enhanced metrology scheme is illustrated in Fig.~\ref{fig1}(c). Under external field driving, the system undergoes a non-equilibrium dissipative phase transition and exhibits a significantly enhanced response near the critical point. As an example, we show the transition from the thermal equilibrium phase to the CTC phase in Fig.~\ref{fig1}(c1). Figures.~\ref{fig1}(c2-c5) present the measured probe transmission in both the time and frequency domains for the thermal equilibrium and CTC phases. Due to the presence of dissipation in the system, the system dynamics can be described using the Lindblad master equation:
\begin{equation}
        \partial_t \hat{\rho} = i [\hat{H},\hat{\rho}] + \mathcal{L}_{R_1}[\hat{\rho}] + \mathcal{L}_{R_2}[\hat{\rho}] + \mathcal{L}_{R_3}[\hat{\rho}]
    \end{equation}
The Lindblad superoperators are given by 
\begin{equation}
       \mathcal{L}_r = (\gamma/2) \sum_i (\hat{\sigma}_i^{r g} \hat{\rho} \hat{\sigma}^{ gr}_i - \{\hat{n}_i^{r},\hat{\rho}\})
\end{equation}
which represent the decay process from the Rydberg state $\left| r \right\rangle$  ($r={R_1,R_2,R_3}$) to the ground state $\left| g \right\rangle$, $\gamma$ represents the decay rate. 

We have simulated the phase transition driven by the microwave field, the simulated results show the predicted increase in oscillation lifetime near criticality, as detailed in the Supplementary Materials. Increasing the microwave drive $\Omega_{\rm{MW}}$ toward the critical value $\Omega^c_{\rm{MW}}$ gradually prolongs oscillation lifetimes, the effect of critical slowing down occurs \cite{wissel1984universal,zhang2024early}. Near the critical point $\Omega^c_{\rm{MW}}$ (corresponding to the critical amplitude $E_c$), the system’s response slows dramatically: decay times diverge, reflecting the system’s increasing “hesitation” to return to equilibrium. In this regime, the system becomes highly sensitive to small perturbations. Once $\Omega_{\rm{MW}} >\Omega^c_{\rm{MW}}$, persistent oscillations emerge, signaling the stability of the time-crystal phase. This mechanism provides a foundation for an enhanced microwave-field sensing methodology based on critical dynamics. In the experiment, we employ a three-photon electromagnetically induced transparency (EIT) protocol to probe the Rydberg atom populations \cite{harris1990nonlinear,petrosyan2011electromagnetically,mohapatra2007coherent}. A schematic of the experimental setup is presented in Fig.~\ref{fig1}(b), see more details in Method sections.

\begin{figure*}
    % Fig3 caption 已更新---20251101
    \centering
    \includegraphics[width=2.08\columnwidth]{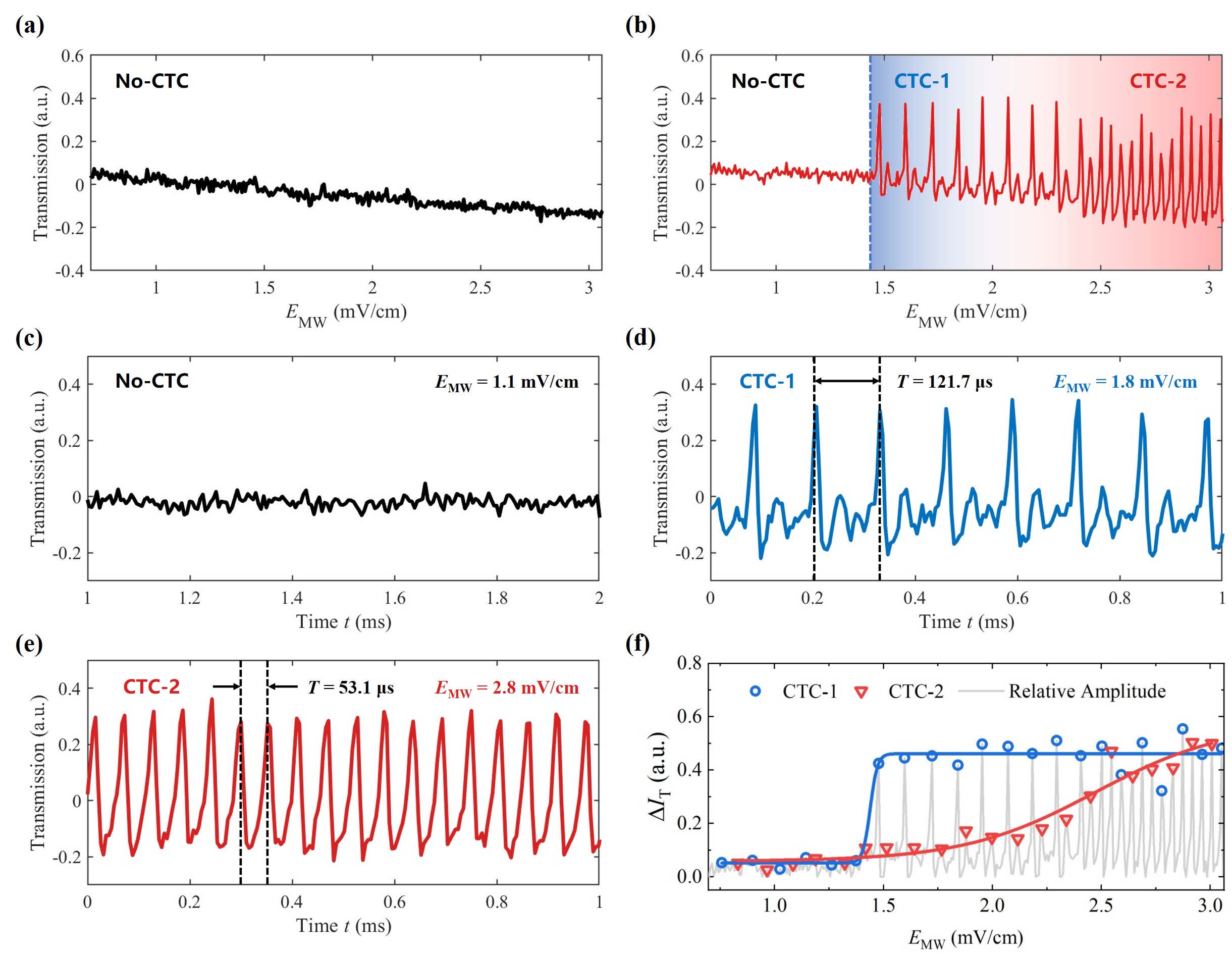}
    \caption{\textbf{Cascaded phase transitions with scanning the microwave amplitude.} (a) The measured transmission $I_{\rm{T}}$ with $E_{\rm{MW}}$ varying from 0.7 mV/cm to 3.06 mV/cm when the RF field is turned off. The system exhibits no evidence of time crystal phase. (b) The recorded transmission $I_{\rm{T}}$ when the RF field is turned on, the system undergoes phase transition from the no-continuous-time-crystalline (no-CTC) phase to the CTC-1 phase and finally to the CTC-2 phase by increasing $E_{\rm{MW}}$. (c)-(e) represent the time-domain response in different phases, where (c) corresponds to no-CTC phase with $E_{\rm{MW}}=1.1$ mV/cm, (d) corresponds to the CTC-1 phase with $E_{\rm{MW}}=1.8$ mV/cm, and (e) corresponds to the CTC-2 phase with $E_{\rm{MW}}=2.8$ mV/cm, respectively. (f) Critical scaling in phase transitions. We extracted the relative oscillation amplitude of the transmission spectrum, as shown by the blue and red data points. The blue and red lines correspond to the critical scaling from the no-CTC phase to the CTC-1 phase and from the CTC-1 phase to the CTC-2 phase, respectively. The fit function is $\Delta I_{\rm{T}} = A/(1 + e^{B(E_{\rm{MW}} - E_0)}) + C$, with $A=-0.41, B=59.63, E_0= 1.44, C=0.46$ (blue line) and $A=-0.52, B=3.31, E_0= 2.47, C=0.58$ (red line).}
    \label{fig3}
\end{figure*}

\subsection*{Multi-parameter phase diagram}
    To experimentally demonstrate criticality-enhanced sensing via transition of a time crystalline phase, we have systematically measured the full phase diagrams to explore the underlying physical processes. The phase diagram of the system relies on three control parameters, namely microwave frequency $f_{\rm{MW}}$, microwave amplitude $E_{\rm MW}$, and laser detuning $\Delta$. We first fix the amplitude to $E_{\rm MW}=9.33~\rm{mV/cm}$ and scan the microwave frequency $f_{\rm{MW}}$  from 6.405 GHz to 6.605 GHz and the detuning $\Delta$ from $-2\pi \times 126.8$ MHz to $2\pi \times 145.8$ MHz, as shown in Fig.~\ref{fig2}(a). We observe an anti-crossing response in the transmission spectrum. The underlying mechanism can be understood as follows: the microwave field strongly couples to the atomic RF-transition between Rydberg states $\ket{R_2}$ and $\ket{R_3}$, forming dressed states that manifest as the observed anti-crossing in the probe transmission spectrum. Interestingly, when the system is driven into a specific parameter regime (marked by the gray areas in the spectrum given in Figs.~\ref{fig2}(b) with frequency $f_{\rm{MW}}$ = 6.455 GHz), it undergoes a non-equilibrium dissipative phase transition where continuous time-translation symmetry is spontaneously broken and a time-crystalline phase emerges. Figure~\ref{fig2}(c) corresponds to the case of a non-equilibrium phase transition by changing $f_{\rm{MW}}$.

    In order to further analyze the multi-parameter dependence of the phase diagram, we also fix the frequency to the resonance point $f_{\rm{MW}}$ = 6.505 GHz while scanning the detuning $\Delta$ from $-2\pi \times 98.6$ MHz to $2\pi \times 93.8$ MHz and the microwave field amplitude $E_{\rm MW}$ from $1.66~\rm{mV/cm}$ to $16.58~\rm{mV/cm}$. The obtained phase diagram is illustrated in Fig.~\ref{fig2}(d). As $E_{\rm MW}$ increases, the probe transmission spectrum splits into two peaks due to the Autler-Townes (AT) effect \cite{sedlacek2012microwave}, accompanied by the emergence of the oscillation regime. Figure.~\ref{fig2}(e) displays the measured spectrum as a function of detuning $\Delta $, while Fig.~\ref{fig2}(f) shows the probe transmission as a function of amplitude $E_{\rm MW}$ at a fixed detuning $\Delta=-2\pi\times37.9 $ MHz. The enhanced response of transmission can be found in Fig.~\ref{fig2}(f), marked by a sharp transition when the spectrum passes through the critical point, where the system’s susceptibility to the perturbations of the microwave field is dramatically amplified. 

    These measured phase diagrams allow us to accurately map the transition boundary between the thermal equilibrium phase and the CTC phase against multiple parameters. In the following, we show that the phase transitions driven by different parameters can be used for sensing. While three parameters are used to identify the phase diagram, for the sake of brevity we focus on quantum sensing for two of them, namely the microwave field amplitude $E_{\rm MW}$ as well as the frequency $f_{\rm{MW}}$.

\begin{figure*}
% 更新时间：2025-11-11
    \centering
    \includegraphics[width=1.9\columnwidth]{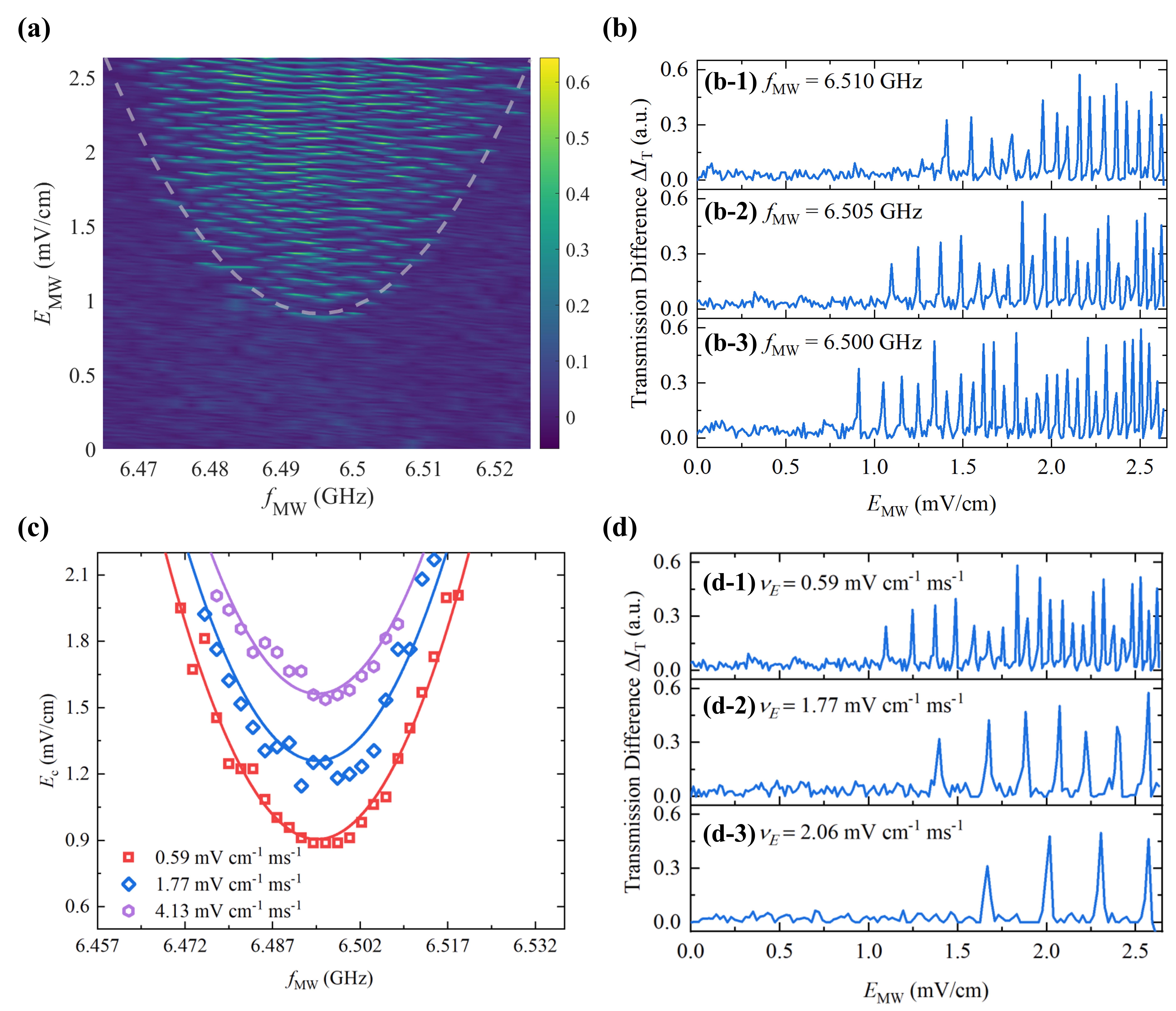}
    \caption{\textbf{Multi-parameter enhanced sensing.} (a) Measured transmission difference with $f_{\rm{MW}}$ sweeping from 6.465 GHz to 6.525 GHz and $E_{\rm{MW}}$ from 0 to 2.63 mV/cm over 5 ms. Color scale indicates the probe transmission intensity $I_{\rm{T}}$. The transmission difference $\Delta I_{\rm{T}}$ is obtained by subtracting a fitted baseline to isolate microwave-induced variations. The CTC phase, identified as the region of sharp spikes, is outlined by a gray dashed line representing the quadratic fit $E_{\rm{MW}}=a_0 f_{\rm{MW}}^2-b_0 f_{\rm{MW}}+c_0$ to its envelope, with $a_0 = 1.93\times 10^{3}$, $b_0 = 2.50\times 10^{4}$, $c_0 = 8.13\times 10^{4}$. The optical detuning $\Delta$ is fixed at approximately $2\pi\times4~\rm{MHz}$ for all measurements. (b) Measured transmission difference $\Delta I_{\rm{T}}$ at fixed microwave frequencies $f_{\rm{MW}}$ = 6.510 GHz (top), 6.505 GHz (middle), and 6.500 GHz (bottom). (c) Phase transition boundaries obtained at microwave electric field scanning rates $\nu_E$ of 0.59 $\rm{mV~cm^{-1}~ms^{-1}}$ (red squares), 1.77 $\rm{mV~cm^{-1}~ms^{-1}}$ (blue diamonds), and 4.13 $\rm{mV~cm^{-1}~ms^{-1}}$ (purple hexagons), respectively. All three datasets are fitted with parabolic functions $E_{\rm{MW}}=a_0 f_{\rm{MW}}^{2}+b_0 f_{\rm{MW}}+c_0$, where the coefficients $a_0$ and $b_0$ are fixed to the values obtained from the global fit in (a), and the red, blue, and purple curves correspond to $c_0=81328.41$, $81328.78$, and $81329.11$, respectively. (d) Evolution of the transmission difference $\Delta I_{\rm{T}}$ with the scanning rate $\nu_E$ at a frequency of $f_{\rm{MW}}=6.505~\rm{GHz}$, as shown for values of 0.59 (top), 1.77 (middle) and 2.06 (bottom) in units of $\rm{mV~cm^{-1}~ms^{-1}}$.}
    \label{fig4}
\end{figure*}

%\subsection*{Microwave driven cascaded criticality}
\subsection*{Amplitude sensing through driven microwave cascaded criticality}
    The phase transition driven by the microwave field amplitude $E_{\rm{MW}}$ is detailed in Fig.~\ref{fig3}. When the RF field is turned off, the system remains in a thermal equilibrium state regardless of the microwave field amplitude, showing no signs of a time-crystalline phase, as shown in Fig.~\ref{fig3}(a). In this case, the transmission $I_{\rm{T}}$ changes linearly with the $E_{\rm{MW}}$, indicating the absence of critical behaviour and the system's response remains within the conventional regime, where perturbations induce proportional changes without enhancement. In contrast, with the RF field applied, the system undergoes a clear non-equilibrium phase transition as $E_{\rm{MW}}$ increases, evolving from a state with no-CTC phase (thermal equilibrium phase) to a first CTC-1 phase, and finally to a second CTC-2 phase with a distinct oscillated frequency, as depicted in Fig.~\ref{fig3}(b). This cascade of transitions represents successive spontaneous breakings of the continuous time-translation symmetry, where the interplay between long-range Rydberg interactions, external driving, and dissipation leads to the emergence of distinct non-equilibrium steady states. In this case, the system's time translation symmetry is broken twice when going across the CTC-1 and CTC-2 phases, displaying enhanced responses characterized by divergent susceptibility near the phase boundaries. In addition, the comb-like structure allows us to measure different microwave field amplitude. The time-domain responses of the system in these different regimes are shown in Figs.~\ref{fig3}(c)-(e). The no-CTC phase exhibits no sustained oscillations [see Fig.~\ref{fig3}(c)], maintaining temporal disorder, while the CTC-1 and CTC-2 phases are characterized by persistent oscillatory dynamics, with the latter demonstrating a modified oscillation pattern, confirming the transition between two separated time-crystalline orders. 
    
    We characterize the microwave-driven criticality properties by recording the amplitude of each peak versus the microwave field amplitude. The measured transmission difference $\Delta I_{\rm{T}}=I_{\rm{T}}-I_{\rm{TF}}$ as given in Fig.~\ref{fig3}(f), where $I_{\rm{TF}}$ represents the transmission with no-RF field. The peaks obtained from the CTC-1 phase are discriminated from the peaks of the CTC-2 phase according to their frequency. The phase transition between the thermal equilibrium phase and the CTC-1 phase exhibits different critical scaling from the transition between the CTC-1 and CTC-2 phases, manifesting as distinct scaling behavior [see the fitted red and blue curves in Fig.~\ref{fig3}(f)] and exhibiting different enhancement factors. This difference arises because the first transition (equilibrium to CTC-1, corresponding to the transition from disordered to ordered) involves the initial emergence of temporal order from a featureless state, governed by the system's approach to a dissipative critical point, while the second transition (CTC-1 to CTC-2, both of which are ordered states) represents a reorganization of existing temporal order, potentially involving different symmetry breaking patterns or interaction pathways. This demonstrates that different types of temporal symmetry breaking provide distinct metrological advantages, with each critical point offering enhanced sensitivity within specific parameter ranges, establishing the foundation for multi-stage quantum enhanced metrology. 
\begin{figure*}
    \centering
    \includegraphics[width=2.0\columnwidth]{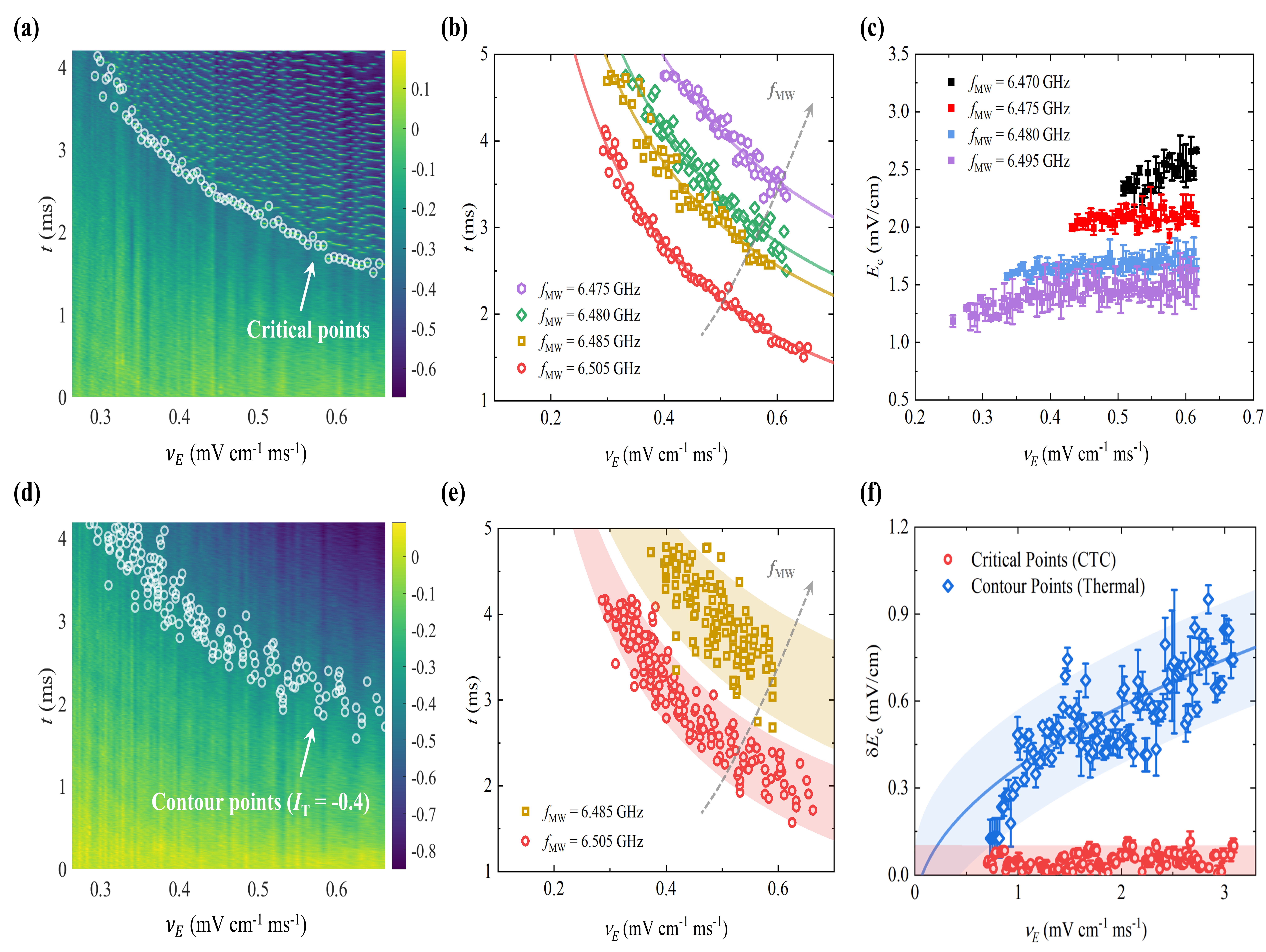}
    \caption{\textbf{Criticality-enhanced sensing and beyond SQL.} (a) Measured phase diagram with varying scanning rates $\nu_{E}$. Color scale indicates the probe transmission intensity $I_{\rm{T}}$. The variation in $\nu_{E}$ shifts the critical point in time domain, forming a phase boundary (white circles). (b) The measured critical points as a function of $\nu_E$ with different $f_{\rm{MW}}$, resulting in distinguishable boundaries. The fit function is $t=A_{\rm{CTC}}/\nu_E + B_{\rm{CTC}}/\sqrt{\nu_E}$, yielding $A_{\rm{CTC}}=1.50$, $B_{\rm{CTC}}=-0.59$ for $f_{\rm{MW}}=6.505$ GHz; $A_{\rm{CTC}}=1.34$, $B_{\rm{CTC}}=0.26$ for $f_{\rm{MW}}=6.485$ GHz; $A_{\rm{CTC}}=1.43$, $B_{\rm{CTC}}=0.35$ for $f_{\rm{MW}}=6.480$ GHz; and $A_{\rm{CTC}}=1.38$, $B_{\rm{CTC}}=0.97$ for $f_{\rm{MW}}=6.475$ GHz. (c) shows the critical amplitude $E_c$ with error $\delta E_c$ within three independent measurements. (d) Measured system response with the RF field off. We fix the position of the contour points (white circles) at a level of -0.4, and defined as the critical amplitude $E_c$ for the case of thermal phase. (e) is the recorded distribution of contour points with $f_{\rm{MW}}=6.485$ GHz and $f_{\rm{MW}}=6.505$ GHz. The fit function is $t=A_{\rm{Th}}/\nu_E + B_{\rm{Th}}/\sqrt{\nu_E}$; the $95\%$ prediction bands are shown as red shading ($f_{\rm{MW}}=6.505~\rm{GHz}$, $A_{\rm{Th}}=1.29$, $B_{\rm{Th}}=-0.03$) and yellow shading ($f_{\rm{MW}}=6.485~\rm{GHz}$, $A_{\rm{Th}}=0.79$, $B_{\rm{Th}}=1.61$). (f) Comparison of measured error $\delta E_c$ between the thermal phase contour points (blue diamonds) and the CTC phase critical points (red circles) at $f_{\rm{MW}}=6.5~\rm{GHz}$. For the thermal system, the error $\delta E_c$ increases with $\nu_E$ and is fitted by $\delta E_c=a_{\rm{Th}}\sqrt{\nu_E}+b_{\rm{Th}}$ (blue solid curve, $a_{\rm{Th}}=0.51$, $b_{\rm{Th}}=-0.13$), with the blue band indicating the $95\%$ prediction interval. In contrast, for the CTC phase, $\delta E_c$ remains nearly constant and consistently lower, as indicated by its $95\%$ prediction band (red shading). Error bars denote the standard deviation from three independent measurements.}

    \label{fig5}
\end{figure*}

%\subsection*{Multi-parameter enhanced sensing}
\subsection*{Microwave frequency sensing}
    The system’s susceptibility diverges at criticality, leading to amplified responses to minute variations [see the multi-parameter-dependent criticality in Fig.~\ref{fig2}(a) and Fig.~\ref{fig2}(d)]. We have investigated the capability of enhanced metrology for both amplitude $E_{\text{MW}}$ and frequency $f_{\rm{MW}}$. We measure the transmission response by scanning the microwave frequency $f_{\rm{MW}}$, and obtain the phase diagram as shown in Fig.~\ref{fig4}(a). The region enclosed by the dashed line indicates the emergence of the CTC phase, characterized by a distinct fingerprinted pattern. 
        
    Near the critical point, sharp changes in the transmission enable highly sensitive detection of tiny variations in both $E_{\text{MW}}$ and $f_{\rm{MW}}$, underscoring the advantages of criticality-based schemes for multi-parameter sensing. The dashed phase boundary line given in Fig.~\ref{fig2}(a) shows a parabolic function of $E_{\text{MW}}$ and $f_{\rm{MW}}$, exhibiting the dependence on two parameters. As shown in Fig.~\ref{fig4}(b), we present the transmission responses measured at microwave frequencies $f_{\rm{MW}}$ = 6.510 GHz, 6.505 GHz, and 6.500 GHz, respectively. In Figs.~\ref{fig4}(b1-b3), each frequency corresponds to a distinct critical microwave amplitude $E_c$, showing the system’s sensitivity to frequency tuning. The observed comb-like structures, combined with the shifting critical points, clearly demonstrate the capability for multi-parameter detection in this critical platform. 
        
    We further measure the phase boundaries under different scanning rates $\nu_E$ with MW electric field $E_{\text{MW}}$ swept from 0.5 to 2.2 mV/cm. The results for $\nu_E$ values of 0.59 $\rm{mV~cm^{-1}~ms^{-1}}$, 1.77 $\rm{mV~cm^{-1}~ms^{-1}}$, and 4.13 $\rm{mV~cm^{-1}~ms^{-1}}$ exhibit three distinctly separated boundary lines, as shown in Fig.~\ref{fig4}(c). This scanning-rate-dependent displacement of the phase boundary is attributed to the non-equilibrium dynamics and finite relaxation timescales of the system during the rapid electric field sweeps. The evolution of the transmission with different $\nu_E$  under the microwave frequency $f_{\rm{MW}}$ = 6.505 GHz is illustrated in Fig.~\ref{fig4}(d). At different scanning rates, the transmission displays different degrees of peak due to the rigid temporal order of time crystal.

%\subsection*{Beyond the SQL}
\subsection*{Quantum enhanced sensitivity}
    To further elucidate the criticality-enhanced metrological behaviour, we systematically investigate the system’s response under different scanning rates $\nu_{\text{E}}$ of the MW electric field. We record the probe transmission $I_{\rm{T}}$ versus time and $\nu_{\text{E}}$ and we can thus probe the system dynamical response near the critical point, as shown in Fig.~\ref{fig5}(a). In this case, we only focus on the critical regime from thermal equilibrium to the CTC-1 phase, where the system exhibits the greatest sensitivity as discussed before. When the system is tuned near the phase boundary [see the white circles in Fig.~\ref{fig5}(a)] between the thermal equilibrium phase and the CTC-1 phase, the system's response becomes strongly nonlinear and sweep-rate dependent, displaying an amplified effect. This results in a sharp and non-linear phase boundary line in Fig.~\ref{fig5}(a) since the system reaches a threshold microwave amplitude $E_c$, which is also dependent on $\nu_{\text{E}}$, see the distinct threshold $E_c$ = 1.08 mV/cm and 1.13 mV/cm in Supplementary Materials. The sharp phase boundary is attributed to the spontaneously emergent temporal order, which arises from the breaking of continuous time-translation symmetry and manifests as a fixed rising edge in the time-domain response. We further measure the phase boundary at different microwave frequencies $f_{\rm{MW}}$, as shown in Fig.~\ref{fig5}(b). The sharpness of the phase boundary enables clear discrimination between different $f_{\rm{MW}}$ values. In Fig.~\ref{fig5}(c), we present the critical amplitude $E_c$ and show the statistical error within multiple measurements.
        
    Without the RF field, the system behaviour is described by the thermal equilibrium phase, the measured result is shown in Fig.~\ref{fig5}(d). We plot a set of contour points and obtain the system transmission, as given by the white circles in Fig.~\ref{fig5}(d).  The contour points in transmission correspond to the same microwave field amplitude $E_{\rm{MW}}$ due to the mapping function $I_{\text{T}}(E_{\rm{MW}})$. The contour points have large fluctuations, corresponding to the worse sensitivity in $E_{\rm{MW}}$. Fig.~\ref{fig5}(e) illustrates the distribution of contour points at selected microwave frequencies (6.505 GHz and 6.485 GHz) for clarity. As $f_{\rm{MW}}$ shifts away from resonance, the distribution of contour points moves in a manner similar to the corresponding critical points in Fig.~\ref{fig5}(b). In this case, the absence of critical fluctuations or divergent susceptibility results in a trivial scaling behaviour, limiting the measurement sensitivity on the microwave field. Since the lowest limit of sensing is inversely proportional to the slope of transmission of the physical quantity (see the Supplementary materials), we can compare the sensitivity between different cases. By comparing the slop of the system response $k = dI_{\text{T}}/dE_{\rm{MW}}$ near the critical point under the cases of criticality and no-criticality (see Supplementary Materials), we estimate the sensitivity difference between these two cases $\alpha =|k_{\rm{CTC}}|/|k_{\rm{Th}}| \sim 323$, indicating a criticality-enhancement for the microwave field intensity sensitivity of approximately 25 dB. This corresponds to the enhancement of Fisher information (FI) in the order of more than $10^5$ by considering nearly same variance between no-criticality and criticality cases.
        
    The sensing error associated with measurements near criticality surpasses the SQL, which scales as $1/\sqrt{N t}$ and represents the fundamental sensitivity barrier for any classical detection scheme that does not utilize quantum resources like entanglement or many-body interactions. In our experiment, the number of Rydberg atoms $N$ is fixed by maintaining a consistent transmission level, allowing us to study the measurement precision as a function of the scanning rate $\nu_{\text{E}}$ (which is proportional to $1/t$). As shown in Fig.~\ref{fig5}(f), the measured error $\delta{E_{\text{MW}}}$ in the thermal equilibrium phase exhibits a $\sqrt{\nu_{\text{E}}}$ scaling, consistent with the SQL. In contrast, under critical conditions, the measurement error becomes extremely low and lies far below the SQL. The measured precision via criticality has reached a floor dominated by the technical noise of the system, obscuring the projected quantum scaling. Consequently, the sensitivity remains constant at a level near the system's noise floor, independent of the scanning rate.

\section*{Discussions and Summary}

In this work, we harness time-translation symmetry breaking in dissipative many-body Rydberg atom systems for creating a CTC, as an exotic non-equilibrium phase of matter with long lasting periodic behavior.  Our experiment has  three main results. First, through characterizing the full phase diagram of the system, we show that a single CTC setup can indeed be used for measuring multiple parameters, a paradigm which is way more complex than single parameter sensing. Second, we observe an interesting cascaded phase transition in which the many-body system goes from a thermal phase to a CTC-1 phase which is then followed by a second transition into a distinct CTC-2 phase. Different features of the CTC-1 and CTC-2 signals can be used for a two-stage tuning of the probe allowing the estimation to always takes place near the CTC-1 phase transition in which the sensitivity is maximal. Third, we reveal a direct connection between the the CTC phase boundary and the scanning rate which allows for achieving precisions beyond the SQL. Our experiment  illustrates the potential of non-equilibrium many-body phases for multiple-parameter sensing and opens pathways toward robust quantum sensors based on temporal order and criticality.

%In the experiment, the observation of distinct separated critical dynamical behaviours across cascaded phase transitions provides valuable insights for advancing more complex quantum-enhanced sensing schemes. This cascaded feature of phase transitions across a broader parameter range suggests a multi-stage criticality which can be used to demonstrate a hierarchical sensing. Moreover, our work highlights a key distinction between metrology based on time-translation symmetry breaking and that relying on spatial or other conventional forms of symmetry breaking \cite{ding2022enhanced}. The breaking of time-translation symmetry manifests as a series of peaks in the linearly scanning spectrum and establishes a robust temporal order, exhibiting a big difference before and after phase transition. While, this effect is not obvious in the case reported in Ref. \cite{ding2022enhanced}.
    
%In summary, we have reported an approach on how to realize the quantum enhanced metrology via exploiting the critical behaviour near a CTC phase transition in a driven-dissipative ensemble of Rydberg atoms. By comparing measurement precision in critical and non-critical regimes, we achieved more than two orders of magnitude improvement in slope, with performance surpassing the SQL. Most importantly, the multi-parameter criticality enhancement has been realized in our experiment. This work not only illustrates the potential of non-equilibrium many-body phases for practical sensing applications but also opens pathways toward robust quantum sensors based on temporal order and criticality.

\section*{Methods}
\subsection*{Experimental setup}
    The specific experimental setup is illustrated in Fig.\ref{fig1}(b). We employed a three-photon excitation scheme to excite Caesium atoms from ground state $\ket{6S_{1/2}}$, via two intermediate states $\ket{6P_{3/2}}$ and $\ket{7S_{1/2}}$, to a thermal equilibrium Rydberg state $\ket{49P_{3/2}}$, and an additional  RF field carried by a pair of electrodes to lift the degeneracy between the substates $\ket{R_1}$ and $\ket{R_2}$, which creates a competition between the substates and consequently drives the Rydberg atoms toward the vicinity of the CTC phase transition critical point. The probe laser [emitted from an 852 nm external-cavity diode laser (ECDL)] and the decoupling laser (emitted from an 1470 nm ECDL) are locked with a saturated absorption spectrum (SAS) and a two-photon EIT spectrum respectively. Coupling laser is emitted from a 780 nm ECDL integrated with a tapered amplifier (TA) to satisfy the high power requirement. The probe, decoupling and coupling lasers have powers of \SI{64}{\micro\watt}, 16.8 mW and 1.5 W, and $1/e^2$ waist radii of approximately \SI{200}{\micro\meter}, \SI{500}{\micro\meter} and \SI{500}{\micro\meter}, respectively, with the corresponding Rabi frequencies as $\Omega_p=2\pi\times$35 MHz, $\Omega_d=2\pi\times$235 MHz and $\Omega_c=2\pi\times$4.3 MHz. The coupling laser frequency is first set to free-running scan to acquire the complete EIT-AT spectrum and its associated oscillation behaviour originating from CTC phase, and is subsequently fixed to monitor the transition spectrum as a function of the target microwave power. Probe laser is aligned by a set of dichroic mirrors to counter-propagate against the decoupling and coupling lasers, coincides with them within a 7 cm atom vapour cell, and is finally collected by a photodetector (Thorlabs PDB450A-AC) to output the transmission spectrum. The RF field, with its frequency and amplitude fixed at 8.2 MHz and 900 mV respectively, is generated utilizing an arbitrary function generator (Rigol DG902PRO AFG) and applied to the atomic vapor cell via a pair of RF electrode plates (120 mm diameter, 3 mm thick, 40 mm separation). We use a signal generator to produce the targeting microwave and an antenna horn to implement it to the atoms. All the equipment and the experimental parameters are connected and centrally controlled by a computer. 

   To achieve a rapid linear sweep of the MW field amplitude $E_{\rm{MW}}$, we utilize the amplitude modulation function of a vector signal generator (VSG, Ceyear 1465F-V), with an external modulation source provided by a rising sawtooth waveform from an arbitrary function generator (Tektronix AFG3235). In our experiments, the modulation depth of the microwave electric field is consistently maintained at $100\%$. The external modulation signal from the AFG also serves as the trigger source for the oscilloscope to ensure synchronized data acquisition. The scanning rate of the microwave electric field, denoted as $\nu_E$, is defined as the ratio of the maximum calibrated electric field amplitude to the duration of a single scanning cycle. By adjusting both the sawtooth frequency from the AFG and the maximum output power of the VSG, different scanning rates $\nu_E$ can be precisely controlled.

\subsection*{Electric Field Amplitude Calibration}
    The microwave electric field amplitude within the $^{85}\rm{Rb}$ vapour cell is calibrated using the AT effect, referenced against a known fine-structure interval. When a microwave source resonantly drives the transition between the $\ket{49P_{3/2}}$ and $\ket{48D_{5/2}}$ Rydberg states at a certain output power $P_0$ from MW source, the probe spectrum's EIT peak splits into a doublet. The splitting frequency, $\Delta f_{\text{AT}} = \mu_0 E_{\rm{MW}}/h$, depends on the transition dipole moment $\mu_0=\langle 48D_{5/2}|e\mathbf{r}|49P_{3/2}\rangle$ and the electric field amplitude $E_{\rm{MW}}$, with $h$ as the Plank constant. The actual value of $\Delta f_{\text{AT}}$ was obtained from the known energy interval $\Delta E_{\text{FS}}$ between the $49P_{1/2}$ and $49P_{3/2}$ fine-structure levels and their proportions measured from the spectrum $k_{\text{AT/FS}}=h\Delta f_{\text{AT}}/\Delta E_{\text{FS}}$. Then we can calibrate the transmission power loss $P_{\rm{loss}}=-57.92$ dBm from the microwave source to the atoms by $P_{\rm{loss}}=P_0-10\lg \left [E_{\rm{MW}}^2 A_{\rm{r}}/(2\eta_0)\right ]$, with $\eta_0$ as the free space intrinsic impedance, $A_{\rm{r}}=w_0 l_{\rm{cell}}$ as the reception aperture of the atoms, $w_0=200~\rm{\mu m}$ as the diameter of the probe laser waist, and $l_{\rm{cell}}=7~\rm{cm}$ as the length of the vapour cell. Incorporating the defined power loss and the relation between $E_{\text{MW}}$ and $P_0$, it yields the actual microwave electric field at the atoms $E_{\text{MW}}^{2}=\left (2\eta_0/A_{\rm{r}}\right) 10^{(P_0+P_{\rm{loss}})/10}$ with an arbitrary microwave source output power $P_0$.

\section*{Acknowledgements}
    We acknowledge funding from the National Key R and D Program of China (Grant No. 2022YFA1404002), the National Natural Science Foundation of China (Grant Nos. T2495253, 62435018,  12274059, 12574528, 1251101297 and W2541020).

\section*{Data Availability}
    All experimental data used in this study are available from the corresponding author upon request.

\section*{Author contributions statement}
    D.-S.D., A.B., B.-B.W., and L.H.Z. conceived the idea and supported this research. B.L., J.R.C., Y.M.  and Q.F.W. conducted the physical experiments. B.L., D.-S.D., and A.B. developed the theoretical model. The manuscript was written by D.-S.D., J.R.C., B.L., and A.B.. All authors contributed to discussions regarding the results and the analysis contained in the manuscript.

\section*{Competing interests}
    The authors declare no competing interests.

\bibliography{ref}

\end{document}

\title{Supplemental Materials: Enhanced multi-parameter metrology in dissipative Rydberg atom time crystals}

\author{Bang Liu$^{1,2,\textcolor{blue}{\star}}$}
\author{Jun-Rong Chen$^{3,\textcolor{blue}{\star}}$}
\author{Yu Ma$^{1,2,\textcolor{blue}{\star}}$}
\author{Qi-Feng Wang$^{1,2,\textcolor{blue}{\star}}$}
\author{Tian-Yu Han$^{1,2}$}
\author{Hao Tian$^{3}$}
\author{Yu-Hua Qian$^{4}$}
\author{Guang-Can Guo$^{1,2}$}
\author{Li-Hua Zhang$^{1,2,\textcolor{blue}{\#}}$}
\author{Bin-Bin Wei$^{5,\textcolor{blue}{\ddagger}}$}
\author{Abolfazl Bayat$^{6,\textcolor{blue}{\S}}$}
\author{Dong-Sheng Ding$^{1,2,\textcolor{blue}{\dagger}}$}
\author{Bao-Sen Shi$^{1,2}$}

\affiliation{$^1$Key Laboratory of Quantum Information, University of Science and Technology of China; Hefei, Anhui 230026, China.}
\affiliation{$^2$Anhui Province Key Laboratory of Quantum Network, University of Science and Technology of China, Hefei 230026, China.}
\affiliation{$^3$School of physics, Harbin Institute of Technology, Harbin, Heilongjiang 150001, China.}
\affiliation{$^4$Institute of Big Data Science and Industry, Shanxi University, Taiyuan, China}
\affiliation{$^5$Institute of system engineering, Tianjin 300161, China.}
\affiliation{$^6$Institute of Fundamental and Frontier Sciences, University of Electronic Science and Technology of China, Chengdu 611731, China}

\date{\today}

\symbolfootnote[1]{B.L., J.R.C., Y.M. and Q.F.W. contribute equally to this work.}
\symbolfootnote[2]{zlhphys@ustc.edu.cn}
\symbolfootnote[3]{weibb.2009@tsinghua.org.cn}
\symbolfootnote[4]{abolfazl.bayat@uestc.edu.cn}
\symbolfootnote[2]{dds@ustc.edu.cn}

\maketitle

\subsection*{Solution of the Master Equation}
    Based on the system's Hamiltonian, we can derive the system's master equation. Considering the thermal motion of atoms, we neglect the correlations between atoms and employ the mean-field approximation to treat the interactions between Rydberg atoms \cite{wu2023observation,liu2024higher,liu2025bifurcation}. In the mean-field approximation the interaction term $V_{ij}$ is replaced by $V_{\rm{MF}} = \chi(\rho _{R_1R_1}(t)+\rho _{R_2R_2}(t)+\rho _{R_3R_3}(t))$, which represents the mean-field shift of Rydberg atoms. The master equation for the system is then given by:
    \begin{align*}
        \dot{\rho}_{gR_1} &=
        \frac{i}{2}\Omega_1 \rho_{gg}
        + \frac{1}{2}(-\gamma - 2i(\Delta-V_{\rm{MF}}))\rho_{gR_1}
        + \frac{i}{2}\Omega_3 \rho_{gR_3}
        - \frac{i}{2}\Omega_1 \rho_{R_1R_1}
        - \frac{i}{2}\Omega_2 \rho_{R_2R_1}, \\[6pt]
        \dot{\rho}_{gR_2} &=
        \frac{i}{2}\Omega_2 \rho_{gg}
        + \frac{1}{2}(-\gamma - 2i\delta - 2i(\Delta-V_{\rm{MF}}))\rho_{gR_2}
        + \frac{i}{2}\Omega_3 \rho_{gR_3}
        - \frac{i}{2}\Omega_1 \rho_{R_1R_2}
        - \frac{i}{2}\Omega_2 \rho_{R_2R_2}, \\[6pt]
        \dot{\rho}_{gR_3} &=
        \frac{i}{2}\Omega_3 \rho_{gR_1}
        + \frac{i}{2}\Omega_3 \rho_{gR_2}
        + \frac{1}{2}(-\gamma - 2i\delta - 2i(\Delta-V_{\rm{MF}}))\rho_{gR_3}
        - \frac{i}{2}\Omega_1 \rho_{R_1R_3}
        - \frac{i}{2}\Omega_2 \rho_{R_2R_3}, \\[6pt]
        \dot{\rho}_{R_1R_1} &=
        -\frac{i}{2}\Omega_1 \rho_{gR_1}
        + \frac{i}{2}\Omega_1 \rho_{R_1g}
        - \gamma \rho_{R_1R_1}
        + \frac{i}{2}\Omega_3 \rho_{R_1R_3}
        - \frac{i}{2}\Omega_3 \rho_{R_3R_1}, \\[6pt]
        \dot{\rho}_{R_1R_2} &=
        -\frac{i}{2}\Omega_1 \rho_{gR_2}
        + \frac{i}{2}\Omega_2 \rho_{R_1g}
        + \frac{1}{2}(-2\gamma - 2i\delta)\rho_{R_1R_2}
        + \frac{i}{2}\Omega_3 \rho_{R_1R_3}
        - \frac{i}{2}\Omega_3 \rho_{R_3R_2}, \\[6pt]
        \dot{\rho}_{R_1R_3} &=
        -\frac{i}{2}\Omega_1 \rho_{gR_3}
        + \frac{i}{2}\Omega_3 \rho_{R_1R_1}
        + \frac{i}{2}\Omega_3 \rho_{R_1R_2}
        + \frac{1}{2}(-2\gamma - 2i\delta)\rho_{R_1R_3}
        - \frac{i}{2}\Omega_3 \rho_{R_3R_3}, \\[6pt]
        \dot{\rho}_{R_2R_2} &=
        -\frac{i}{2}\Omega_2 \rho_{gR_2}
        + \frac{i}{2}\Omega_2 \rho_{R_2g}
        - \gamma \rho_{R_2R_2}
        + \frac{i}{2}\Omega_3 \rho_{R_2R_3}
        - \frac{i}{2}\Omega_3 \rho_{R_3R_2}, \\[6pt]
        \dot{\rho}_{R_2R_3} &=
        -\frac{i}{2}\Omega_2 \rho_{gR_3}
        + \frac{i}{2}\Omega_3 \rho_{R_2R_1}
        + \frac{i}{2}\Omega_3 \rho_{R_2R_2}
        - \gamma \rho_{R_2R_3}
        - \frac{i}{2}\Omega_3 \rho_{R_3R_3}, \\[6pt]
        \dot{\rho}_{R_3R_3} &=
        -\frac{i}{2}\Omega_3 \rho_{R_1R_3}
        - \frac{i}{2}\Omega_3 \rho_{R_2R_3}
        + \frac{i}{2}\Omega_3 \rho_{R_3R_1}
        + \frac{i}{2}\Omega_3 \rho_{R_3R_2}
        - \gamma \rho_{R_3R_3}.
    \end{align*}

\begin{figure*}
        \centering
        \includegraphics[width=1\columnwidth]{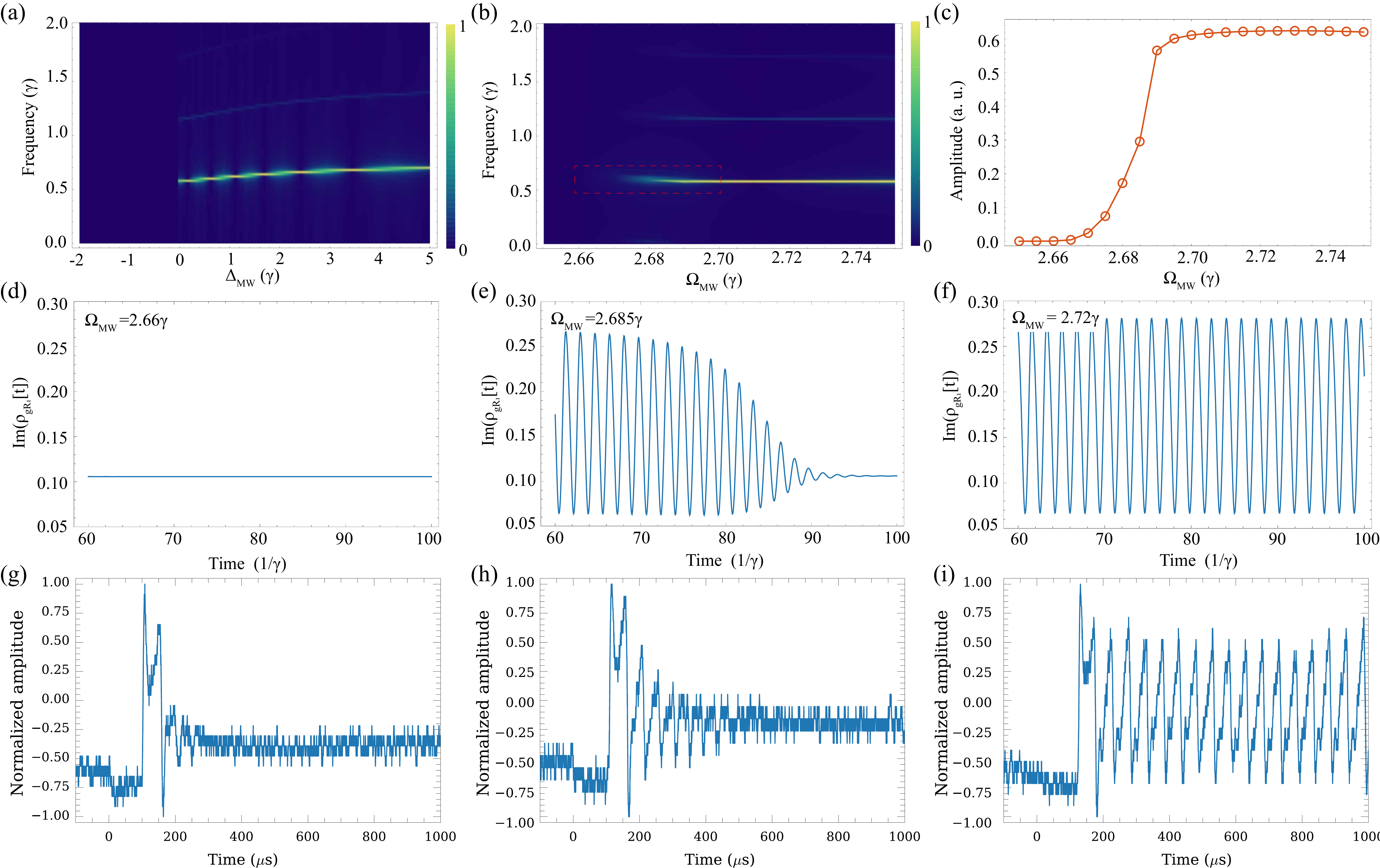}
        \caption{\textbf{Results of theoretical simulation.} (a) Phase diagram of the Fourier spectrum of system response versus microwave detuning $\Delta_{\rm{MW}}$. (b) Phase diagram of the Fourier spectrum of system response versus the Rabi frequency of microwave $\Omega_{\rm{MW}}$. The colorbars in (a) and (b) represent the normalized intensity of the Fourier spectrum. When changing either the microwave detuning or the Rabi frequency, the system undergoes a phase transition from thermal-equilibrium state to a time crystal. (c) When the system undergoes a phase transition, the Fourier spectrum intensity of the system response varies with the Rabi frequency of microwave, corresponding to the red dashed box region in (b). (d), (e), and (f) represent the system's time-domain responses when $\Omega_{\rm{MW}}=2.66\gamma$, $2.685\gamma$, and $2.72\gamma$, respectively. (g), (h), and (i) show the experimentally measured time-domain probe transmission corresponding to the theoretical results in (d–f). The experimental traces reproduce the theoretically predicted increase in oscillation lifetime and the emergence of persistent oscillations.}
        \label{FigS1}
    \end{figure*}

    %where $V_{\rm{MF}} = \chi(\rho _{R_1R_1}(t)+\rho _{R_2R_2}(t)+\rho _{R_3R_3}(t))$ is the mean-field shift of Rydberg atoms. 
    We numerically solved the master equation to simulate the system dynamics under various parameters, as shown in Fig.~\ref{FigS1}.

    When changing the frequency and intensity of the microwave, we observe the system undergoes a phase transition from thermal equilibrium state to a time crystal state. Figure \ref{FigS1}(a) and (b) show the phase diagrams constructed from the Fourier spectra of the system response, plotted as a function of microwave detuning and Rabi frequency. When the system is in the time crystal state, the Fourier spectrum of the system response exhibits bright peaks. During phase transitions, the system exhibits critical behavior near the phase transition point. We scanned the microwave intensity near the critical point and obtained the Fourier spectrum amplitude of the system response, as shown in Fig.~\ref{FigS1}(c). As the microwave Rabi frequency increases (corresponding to an increase in the microwave electric field strength in the experiment), the amplitude of the system response gradually increases, then stabilizes, exhibiting a nonlinear increase. Figure \ref{FigS1}(d-f) illustrate the time-domain response of the system with Rabi frequency $\Omega_{\rm{MW}}=2.66\gamma$, $2.685\gamma$, and $2.72\gamma$, respectively. Here, the imaginary part of the coherence $\rm{Im}[\rho_{gR_1}]$ corresponds to the probe light transmission measured in the experiment. We find that as the microwave Rabi frequency increases, the lifetime of the oscillations is progressively extended. Eventually, the system transitions into a stable time crystal phase, where oscillations persist indefinitely. In order to verify this, we experimentally measured the time-resolved probe transmission to verify the theoretical simulations. Figure \ref{FigS1}(g–i) show the corresponding experimental traces when a microwave field resonant with the Rydberg transition is switched on at time $t$ = 0 $\mu s$. For Fig.~\ref{FigS1}(g), (h), and (i), the corresponding microwave electric field amplitudes are 2.18, 2.65, and 2.97~mV/cm, respectively. These three drive strengths correspond to the cases displayed in Figs.~\ref{FigS1}(d–f). The measured dynamics reproduce the theoretically predicted increase in oscillation lifetime and the emergence of persistent oscillations at higher microwave power, thereby confirming the whole process of the formation of a time-crystal phase. 

\subsection*{Experimental sensing and its connection to estimation theory }

    \begin{figure*}
        \centering
        \includegraphics[width=1\columnwidth]{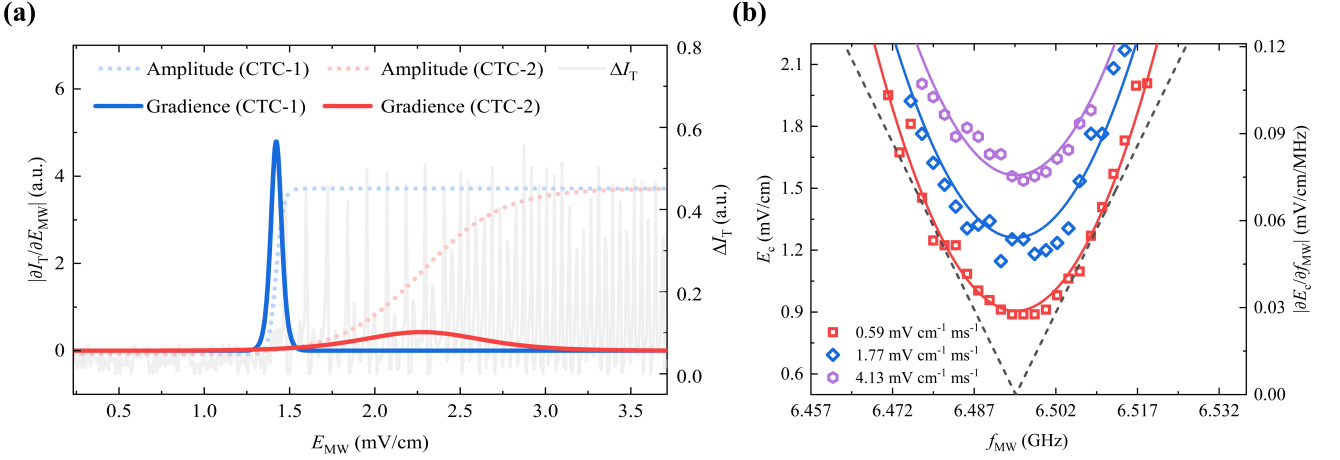}
        \caption{\textbf{Gradients of amplitude signals and phase boundary.} (a) Gradients (absolute slope, $|\partial I_{\rm{T}}/\partial E_{\rm{MW}}|$) of the CTC-1 and CTC-2 amplitude signals as a function of the microwave field amplitude $E_{\rm{MW}}$ (blue and red solid lines, respectively), derived from their corresponding fitting functions (dotted lines). The gradient of CTC-1 exhibits a sharp, high-magnitude peak confined to a narrow region, indicative of an abrupt change. In contrast, the gradient of CTC-2 remains non-zero across a substantially broader range of $E_{\rm{MW}}$, albeit with a consistently lower magnitude, reflecting a more gradual variation. (b) Slope of the phase transition boundary $|\partial E_{\rm{c}} / \partial f_{\rm{MW}}|$ obtained under a slow scanning rate of $\nu_E = 0.59~ \rm{mV~cm^{-1}~ms^{-1}}$ (gray dashed line), plotted together with the corresponding phase boundary (red squares and fitted curve, reproduced from Fig. 4 (c) in the main text). The slope quantifies the resolution of the critical field $E_{\rm{c}}$ to variations in microwave frequency $f_{\rm{MW}}$.}
        \label{Gradience}
    \end{figure*}

For any given quantum probe $\rho(\theta)$, where $\theta$ is the unknown parameter to be estimated, two key elements for accomplishing an estimation task are: (i) a measurement observable $\hat{\Pi}$ which is determined by projectors $\{ \hat{\Pi}_\mu\}$ such that $\hat{\Pi}=\sum_\mu \alpha_\mu\hat{\Pi}_\mu$ (with $\alpha_\mu$ being the experimental measured outcome); and (ii) an estimator which maps the measured data into an  estimation for the unknown parameter. Each measurement outcome $\mu$ takes place with probability $p_\mu(\theta)=\rm{Tr}[\rho(\theta)\hat{\Pi}_k]$. 
In parameter estimation, the Cramér-Rao inequality sets a lower limit on the statistical estimation error from $m$ independent experiments~\cite{lehmann1998theory}
\begin{equation}
    \begin{aligned}
\delta \theta\ge \frac{1}{\sqrt{m  F(\theta)}}
    \end{aligned}
    \end{equation}
Here, $\delta \theta$ is the standard deviation of the estimator and $F(\theta)$ is the Fisher Information (FI), defined as
\begin{equation}
    \begin{aligned}
F(\theta) = \sum_{\mu} \frac{1}{p_\mu(\theta)} \left( \frac{\partial p_\mu(\theta)}{\partial \theta} \right)^2.
    \end{aligned}
    \end{equation}
In practice, however, the measurement setup may not have the resolution to access individual measurement outcomes and instead can only measure the average of the observable operator, namely $I_T=\rm{Tr}[\rho(\theta)\hat{\Pi}]$. In this case, one can extract a lower bound for the Fisher information through error propagation as
\begin{equation}\label{eq:FI_error_prop}
    F(\theta) \ge \frac{(I_{T}^\prime)^2}{m^2(\delta I_T)^2}
\end{equation}
where $I_{T}^\prime$ is the derivative of $I_T$ with respect to $\theta$, namely $I_{T}^\prime=\rm{Tr} [\partial_\theta \rho(\theta) \hat{\Pi}]$ and $\delta I_T$ is the experimental fluctuation in the signal $I_T(\theta)$. This clearly shows that by increasing the derivative $I_{T}^\prime$ the Fisher information increases and thus the estimation precision is improved.

In our protocol, the measurement is always the transmission signal which can be considered as 
\begin{equation}
    I_T(\theta)=-2{\rm Im} {(\rho_{gR_1}(\theta))}={\rm Tr}{[\rho(\theta)\sigma_y]}.
\end{equation}
Here, the experimental observable $\hat{\Pi}$ is  the Pauli matrix $\sigma_y=-i\ket{g}\bra{R_1}+i\ket{R_1}\bra{g}$. We note that the transmission signal which we measure only access to the average of $\sigma_y$ and does not have the resolution of specifying the individual eigenvectors of $\sigma_y$. Therefore, one can extract a lower bound for the Fisher information through the above Eq.~(~\ref{eq:FI_error_prop}).

In order to fulfil the estimation procedure, one has to specify the estimator too. Since our measurement does not have the resolution of accessing individual eigenvectors of the observable, i.e. $\sigma_y$, conventional Bayesian analysis cannot be applied. Instead one can directly rely on the fitting functions of $I_{\rm T}$ to estimate the parameter of interest. To clarify this, in Fig.~\ref{Gradience}(a), we show that the fitting functions of $I_{\rm T}$ in both the CTC1 and CTC2 phases can be used as the estimator from which one can extract the value of $E_{\rm MW}$. By comparing the gradients (slopes) $\partial I_{\rm{T}}/\partial E_{\rm{MW}}$ for the CTC-1 and CTC-2 signals plotted in Fig.~\ref{Gradience}(a), one can see the behavior of the FI, see Eq.~(\ref{eq:FI_error_prop}). Indeed, a distinct behavior becomes evident for CTC1 and CTC2 signals. The gradient extracted from CTC-1 signal is significantly large around the phase transition point but takes very small value otherwise, allowing for very precise estimation at the transition point and poor estimation elsewhere. However, the gradient obtained from the CTC-2 measurement is non-zero over a wide range of amplitudes $E_{\rm MW}$, though its value is sensibly lower than the gradient of CTC-1 around the transition point. Consequently, for estimating $E_{\rm MW}$, one can first employ the broadly responsive CTC-2 signal to obtain a rough estimate of the amplitude. The probe can then be tuned to the vicinity of this estimated value, where the system undergoes the phase transition from thermal equilibrium to the CTC-1 phase. By subsequently measuring the sharply varying CTC-1 signal within this narrow transition region, a highly precise estimation of $E_{\rm MW}$ can be achieved. The same analysis can also be done for the the Microwave frequency. Following the results, presented in Fig. 2(c) of the main text, one can use the the fitting function of $E_c$, which is directly computed from $T_{\rm T}$, to estimate the frequency. An estimation of the FI can also be computed from $\partial E_c/\partial f_{\rm MW}$, see Eq.~(\ref{eq:FI_error_prop}) in which $T_{\rm T}$ is replaced by $E_c$.

\subsection*{Frequency Resolution Estimation}
    \begin{figure*}
        \centering
        \includegraphics[width=1\columnwidth]{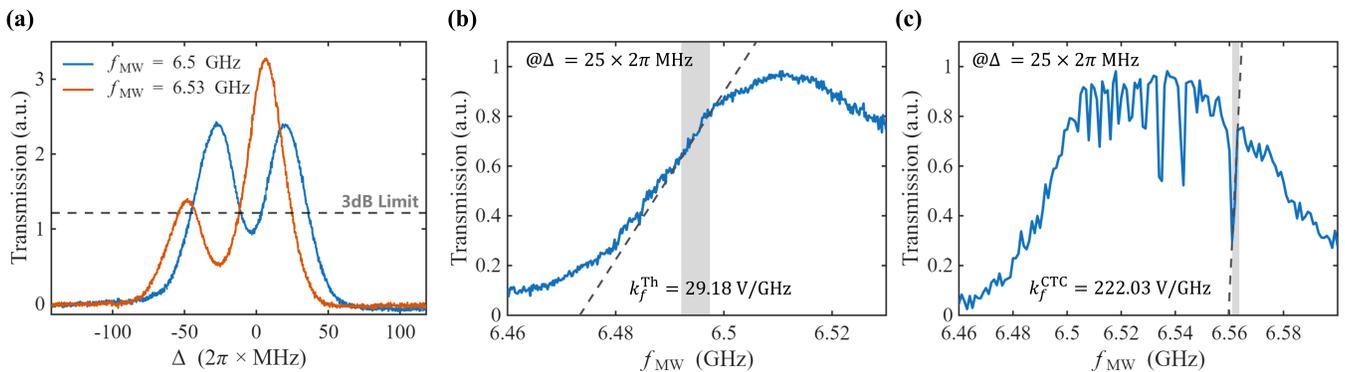}
        \caption{\textbf{Spectroscopic responses and    transduction enhancement across phases.} (a) AT splitting spectra under resonant (blue curve) and off-resonant (orange curve) MW fields. (b) Probe transmission (normalized) for the thermal equilibrium system as a function of MW frequency $f_{\rm{MW}}$ (transduction curve). The gray shading around 6.495 GHz indicates the region of steepest response, with a linear fit (dashed line) giving a slope $k_f^{\rm{Th}}=29.18$ V/GHz. (c) The corresponding transduction curve for the CTC system under an applied RF field exhibits a dramatically sharpened response. The maximum slope near 6.56 GHz reaches $k_f^{\rm{CTC}}=222.03$ V/GHz. The enhanced slope translates to an eightfold improvement in frequency resolution, from $\delta f\approx 1.4$ MHz (b) to $\delta f\approx 180$ kHz (c), under a fixed noise amplitude of approximately 0.04 V.}
        \label{SM_Fig1}
    \end{figure*}
    
    When coupled with both probe and coupling laser, the Rydberg system in thermal equilibrium phase exhibits a resonant enhancement in the probe transmission, manifesting the electromagnetically induced transparency (EIT) effect \cite{harris1990nonlinear,petrosyan2011electromagnetically,mohapatra2007coherent}. The introduction of a resonant microwave (MW) field results in Autler–Townes (AT) splitting of the EIT spectrum \cite{sedlacek2012microwave}. When the MW frequency ($f_{\rm MW}$) deviates from resonance, the resulting AT doublet becomes asymmetric in intensity and shifts in spectral position, as shown in Fig.~\ref{SM_Fig1}(a). 

\begin{figure*}
        \centering
        \includegraphics[width=1\columnwidth]{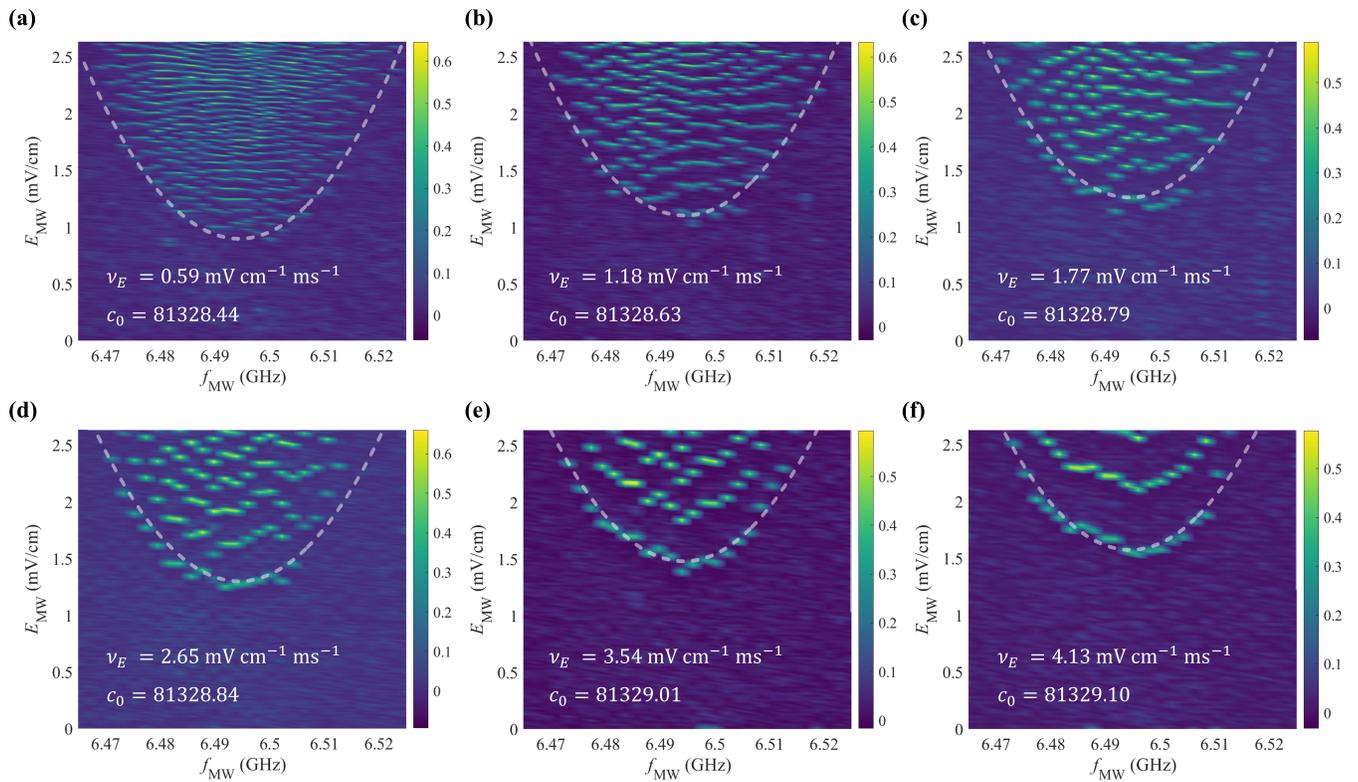}
        \caption{\textbf{Experimental results of the CTC phase diagram under different MW electric field intensity scanning rates $\nu_E$.} The scanning rate $\nu_E$ is adjusted over several $\rm{mV~cm^{-1}~ms^{-1}}$ via varying the scanning frequency $f_{\rm{sc}}$. Values of the scanning rate $\nu_E$ for panels (a) to (f) are achieved at scanning frequencies corresponding to 200, 400, 600, 900, 1200 and 1400 Hz, respectively, with a maximum MW electric field intensity fixed to 2.95 mV/cm. As $\nu_E$ increases, the parabolic CTC phase boundary shifts slightly, with the critical electric field required for the phase transition increasing accordingly. The white dashed curves in each panel represent fits to the quadratic function $E_{\rm{MW}}=a_0 f_{\rm{MW}}^{2}+b_0 f_{\rm{MW}}+c_0$, with fixed coefficients $a_0 = 1928.2$ and $b_0 = -25045.1$, while the intercept $c_0$ differs for each scan rate (as indicated in the corresponding panel). The systematic variation in $c_0$ with $\nu_E$ clearly demonstrates the slight shift of the phase boundary as the scanning rate increases.}
        \label{SM_Fig2}
    \end{figure*}

    Crucially, by monitoring the probe transmission $I_{\rm{T}}$ at a fixed laser detuning while sweeping the microwave frequency $f_{\rm{MW}}$, we obtain a transduction curve that represents the dependence of $I_{\rm{T}}$ on $f_{\rm{MW}}$. The frequency resolution of a quantum sensing protocol is fundamentally governed by the slope of this curve, defined as $k_f=\partial I_{\rm{T}}/\partial f_{\rm{MW}}$. Optimal frequency discrimination occurs at the operating point where $k_f$ is maximized, which corresponds to the highest resolution to frequency variations. For the system in the thermal equilibrium phase [as shown in Fig.~\ref{SM_Fig2}(b)], linear fitting across the steepest response region around 6.495 GHz (gray shading) gives $k_f^{\rm{Th}} \approx 29.18$ V/GHz. In contrast, when driven into the CTC phase by an applied RF field, as exhibited in Fig.~\ref{SM_Fig2}(c), the transduction characteristics are markedly altered. The maximum slope in this regime, located near 6.56 GHz, reaches $k_f^{\rm{CTC}} \approx 222.03$ V/GHz, reflecting a significantly steeper frequency-to-transmission response. Consequently, assuming an experimentally observed noise amplitude of approximately $\delta I_{\rm{T}}=0.04$ V in the transmission signal for both configurations, the attainable frequency resolution $\delta f$ can be directly inferred. For the system in thermal equilibrium phase, $\delta f$ is estimated to be around 1.4 MHz, whereas for the system in CTC phase it reduces to roughly 180 kHz, corresponding to an eightfold enhancement in resolution. This result clearly demonstrates the metrological advantage afforded by operating near the temporal criticality of the CTC phase. Furthermore, the frequency of this maximum‑slope operating point can be further tuned via optical detuning, enabling flexible adaptation of the operational range and system response.

    By varying the microwave scanning frequency $f_{\rm{sc}}$ supplied by the arbitrary function generator, we further measured the CTC phase diagrams under different MW electric field intensity scanning rates $\nu_E$, as shown in Fig.~\ref{SM_Fig2}. The results reveal that the position of the CTC phase boundary shifts slightly with increasing $\nu_E$, where a higher scanning rate necessitates a greater critical electric field intensity $E_c$ to drive the phase transition in the Rydberg atom ensemble. The parabolic shape of these phase boundaries, consistent across all scanning rates, aligns with the behaviour illustrated in Fig. 4(c) of the main text. Furthermore, as the scanning period decreases with increasing $f_{\rm{sc}}$, while the characteristic oscillation period of the CTC phase remains approximately constant, the oscillations become progressively sparser within the shrinking CTC phase region [compare, for example, Fig.~\ref{SM_Fig2}(a) with Fig.~\ref{SM_Fig2}(f)]. It is noteworthy, however, that whenever the system resides within the CTC phase region, the corresponding probe transmission signal exhibits the hallmark oscillations of the time crystal phase. Consequently, the boundary defined by the outermost oscillation peak of the CTC phase region remains a robust indicator for distinguishing between the thermal equilibrium phase and the CTC phase.

\subsection*{Sensitivity Enhancement Estimation}
    \begin{figure*}
        \centering
        \includegraphics[width=1\columnwidth]{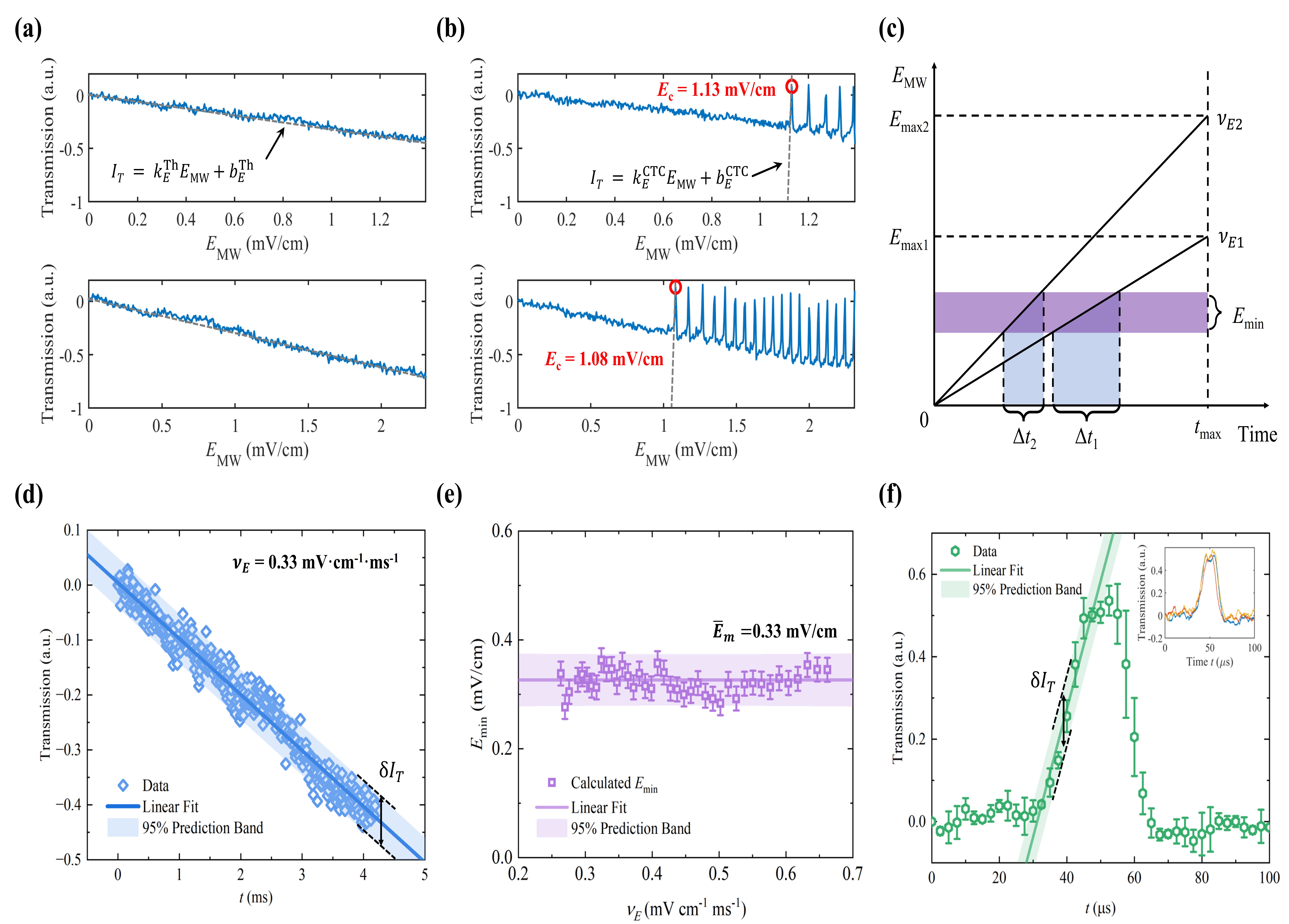}
        \caption{\textbf{Response of the probe transmission to the microwave electric field $E_{\rm{MW}}$ and estimation of the minimum resolvable field under different scanning rate $\nu_E$.} (a) Probe transmission response of the thermal equilibrium system to $E_{\rm{MW}}$ at scanning rate of 0.33 (upper panel) and 0.55 (lower panel) $\rm{mV~cm^{-1}~ms^{-1}}$. Linear fits $I_{\rm{T}} = k^{\rm{Th}}_E E_{\rm{MW}} + b^{\rm{Th}}_E$ (gray dashed lines) yield slopes $ k^{\rm{Th}}_E $ of 0.096 and 0.171 $\rm{V/(mV~cm^{-1})}$, respectively. (b)  Probe transmission response of the CTC system to $E_{\rm{MW}}$ at $\nu_E = $ 0.33 (upper panel) and 0.55 (lower panel) $\rm{mV~cm^{-1}~ms^{-1}}$, with critical field $E_c$ (red circles) of 1.13 and 1.08 mV/cm, respectively. Linear fits $I_{\rm{T}} = k^{\rm{CTC}}_E E_{\rm{MW}} + b^{\rm{CTC}}_E$ (gray dashed lines) to the rising edge of the initial CTC peak give slopes $k^{\rm{CTC}}_E$ of 12.7 and 12.5 $\rm{V/(mV~cm^{-1})}$. (c) Schematic of $E_{\rm{MW}}$ versus time for different $\nu_E$ at a fixed microwave scanning frequency ($f_{\rm{sc}} = 200 ~\rm{Hz}$). For the same measurement duration, the thermal system exhibits a consistent minimum resolvable field (purple shading) but different noise-equivalent durations (blue shadings). (d) Experimental data and linear fit for the thermal equilibrium system, with slope $k^{\rm{Th}}_t$ = -0.102 V/ms and intercept $b^{\rm{Th}}_t = 4.46\times10^{-2}$ V at $\nu_E = 0.33~\rm{mV~cm^{-1}~ms^{-1}}$. (e) Minimum resolvable electric field $E_{\rm{min}}$ of the thermal system across different $\nu_E$ values. The purple solid line indicates the mean of all data (approximately 0.33 mV/cm), with error bars representing the standard deviation. (f) Measured lineshape of the first CTC peak at the critical point. Error bars originate from three independent measurements (see inset). A linear fit to the rising edge gives a slope $k^{\rm{CTC}}_t$ of approximately 31.05 V/ms.}
        \label{SM_Fig3}
    \end{figure*}
    
    The sensitivity of our Rydberg system is fundamentally determined by the product of the minimum resolvable MW electric field intensity $E_{\rm{min}}$ and the corresponding measurement time $t_{\rm m}$. In our experiments, the duration of a single measurement is determined solely by the microwave scanning frequency $f_{\rm {sc}}$. The scanning rate of the microwave electric field $\nu_E$ can also be tuned within a certain range by varying the maximum field strength $E_{\rm{max}}$. Fig.~\ref{SM_Fig3} (a) illustrates the variation of the probe transmission $I_{\rm T}$ for the system in the thermal equilibrium phase at $\nu_E$ values of 0.33 (upper panel) and 0.55 (lower panel) in units of $\rm{mV~cm^{-1}~ms^{-1}}$, both exhibiting high linearity characterized by the slope $k^{\rm{Th}}_{E}$. The corresponding probe transmission $I_{\rm T}$ for the system in the CTC phase at the same $\nu_E$ is presented in Fig.~\ref{SM_Fig3} (b) (upper panel: 0.33 $\rm{mV~cm^{-1}~ms^{-1}}$, lower panel: 0.55 $\rm{mV~cm^{-1}~ms^{-1}}$), where a clear phase transition from the thermal equilibrium to the CTC phase is observed upon increasing $E_{\rm{MW}}$. The corresponding critical electric field strengths $E_c$, marked by red circles, are measured to be 1.13 and 1.08 mV/cm, respectively. It is important to note that the apparent lower $E_c$ under a higher scanning rate $\nu_E$ does not contradict the trend established in Fig.~\ref{SM_Fig2}, considering the inherent uncertainty in determining the phase boundary with the limited variation range of $\nu_E$. Owing to the sharp signal peaks in the CTC phase, a linear fit is employed to approximate the change in probe transmission near the critical point. The detection sensitivity $S$ is governed by the rate of change of the probe transmission $I_{\rm T}$ with respect to the measured electric field $E_{\rm{MW}}$, and a steeper slope $k_E$ implies higher sensitivity to $E_{\rm{MW}}$. Therefore,  $k_E^{\rm{Th}}$ and $k_E^{\rm{CTC}}$ are used to estimate the system's detection sensitivity before and after the criticality-enhancement, respectively.

    \begin{figure*}
        \centering
        \includegraphics[width=1\columnwidth]{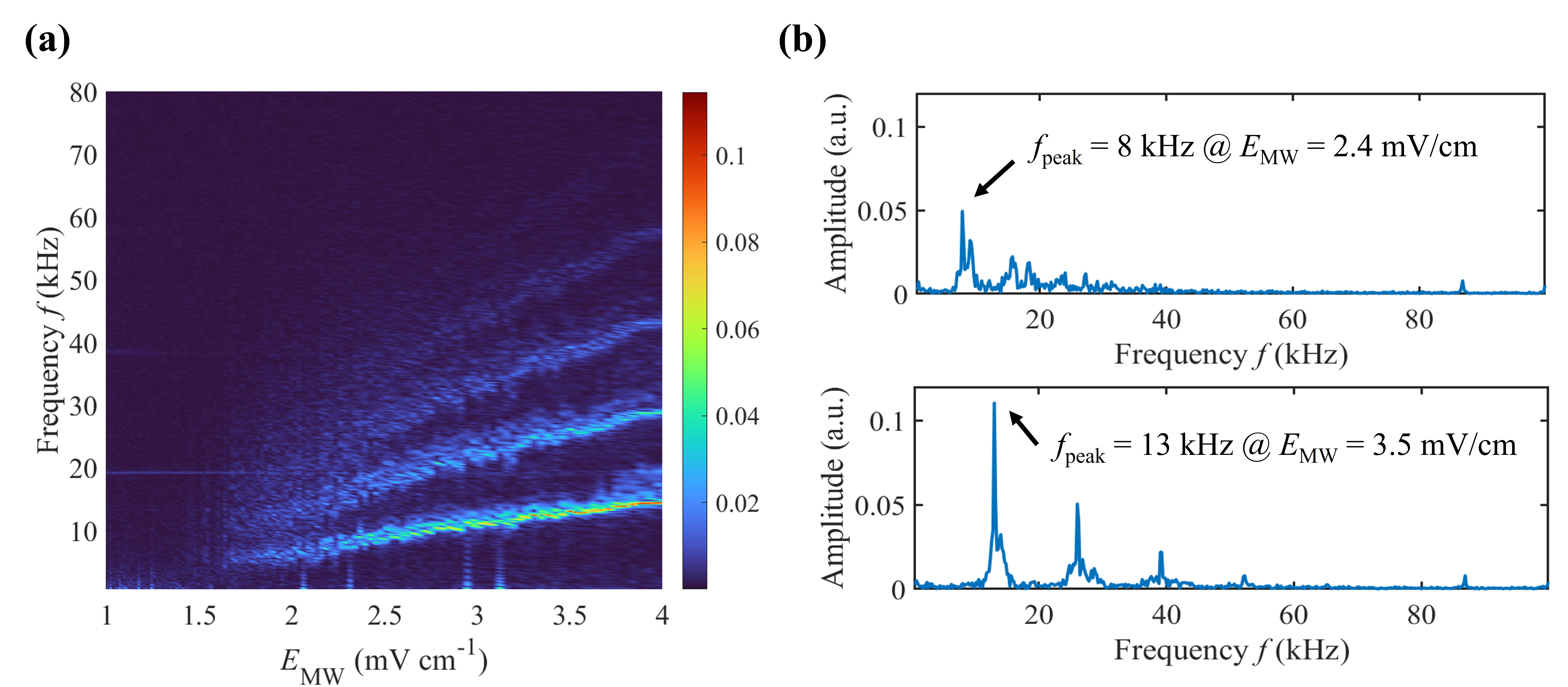}
        \caption{\textbf{Evolution of the FFT spectrum during the time crystal phase transition.} (a) At the microwave resonance frequency, as the microwave electric field intensity $E_{\rm{MW}}$ increases (1–2 mV/cm), the FFT spectrum system's output signal evolves from a featureless background to one exhibiting a fundamental frequency ($\sim$6 kHz) and its higher-order harmonics.  With a further increase in $E_{\rm{MW}}$ (2–4 mV/cm), both the fundamental and harmonic frequencies shift to higher values, accompanied by a simultaneous growth in the signal intensity and enhanced robustness of the time crystal phase. (b) FFT spectra of the time crystal signal recorded at $E_{\rm{MW}} = 2.4$ (upper panel) and 3.5 mV/cm (lower panel), respectively. Note that the peak near 87 kHz arises from the laser modulation for frequency stabilization and is unrelated to the time crystal.}
        \label{SM_Fig4}
    \end{figure*}

    The MW scanning protocols corresponding to Fig.~\ref{SM_Fig3} (a) and (b) are depicted in Fig.~\ref{SM_Fig3} (c), where the microwave scanning frequency $f_{\rm{sc}}$ is held constant, and the scanning rate $\nu_E$ is adjusted by varying the maximum electric field intensity $E_{\rm{max}}$. We define the minimum resolvable electric field $E_{\rm{min]}}$ as the noise-equivalent field $E_{\rm{noise}}$ at a unity signal-to-noise ratio, which equals the product of the noise-equivalent duration $\Delta t$ and the current scanning rate $\nu_E$. Here, $\Delta t = \delta I_{\rm{T}} / k_t$, with $\delta I_{\rm{T}}$ representing the uncertainty in the probe transmission induced by the technical noise and $k_t$ denoting the time derivative of the output signal. Consequently, the minimum resolvable electric field $E_{\rm{min
    }}$ is expressed as
    \begin{align}
        E_{\rm{min}}=\frac{\delta I_{\rm{T}}}{k_E} = \frac{\delta I_{\rm{T}}}{k_t} \nu_E
        \label{fun1}
    \end{align}
    
    For the thermal equilibrium system, the temporal evolution of its output signal at $\nu_E = 0.33~\rm{mV~cm^{-1}~ms^{-1}}$ is illustrated in Fig.~\ref{SM_Fig3} (d). The pronounced linearity stems from the linear MW sweep and the system's linear response to $E_{\rm{MW}}$, characterized by a slope $k^{\rm{Th}}_t$ that relates to the slope $k^{\rm{Th}}_E$ through $k^{\rm{Th}}_t = k^{\rm{Th}}_E\nu_E$. The technical noise floor, estimated from the $95\%$ prediction band of the linear fit, is approximately 0.1 V and remains consistent across different values of $\nu_{E}$, yielding a minimum resolvable electric field of about 0.34 mV/cm for the thermal equilibrium system. In practice, the same methodology can be applied to determine $E_{\rm{min}}$ across different $\nu_E$ under the same scanning frequency $f_{\rm{sc}}$. The resulting $E_{\rm{min}}$ values are expected to be approximately consistent for the same measurement time $t_{\rm{m}}$, as shown in Fig.~\ref{SM_Fig3} (e). The average minimum resolvable electric field for the thermal equilibrium system at this scanning frequency is thus estimated as 0.33 mV/cm. 

    For the system in the CTC phase, the line shape of the first CTC peak at the critical point remains nearly invariant across different scanning rates $\nu_E$ investigated. This consistency enables the average rising-edge slope $k^{\rm{CTC}}_t$ to be determined from a linear fit to the initial CTC peaks acquired at multiple $\nu_E$ values, as illustrated in Fig.~\ref{SM_Fig3} (f). At a specific scanning rate of $\nu_E = 0.33~\rm{mV~cm^{-1}~ms^{-1}}$, the minimum resolvable electric field for the CTC system is estimated utilizing Eq.~\ref{fun1} to be approximately 1.1 $\rm{\mu V/cm}$, representing an enhancement of roughly 25 dB compared to the thermal equilibrium system. Furthermore, with a fixed measurement time of $t_m = 5~\rm{ms}$ (far exceeding the Rydberg coherence time), the MW electric field sensitivity is given by
    \begin{align}
        S_{E}=E_{\rm{min}}\sqrt{t_{\rm{m}}} = \frac{\delta I_{\rm{T}}}{k_t} \nu_E \sqrt{t_{\rm{m}}}
        \label{fun2}
    \end{align}
    Based on this expression, the sensitivity for the CTC system at $\nu_E = 0.33~\rm{mV~cm^{-1}~ms^{-1}}$ is found to be 75.5 $\rm{nV~cm^{-1}~Hz^{-1/2}}$, exhibiting a 25 dB enhancement over the sensitivity of 23.1 $\rm{\mu V~cm^{-1}~Hz^{-1/2}}$ for the thermal equilibrium system.
    
    We further experimentally investigated the evolution of the output signal spectrum during the formation of the CTC phase as a function of the microwave electric field intensity $E_{\rm{MW}}$, with the results presented in Fig.~\ref{SM_Fig4}. When the field intensity $E_{\rm{MW}}$ is below the critical field $E_{\rm{c}}$, the output spectrum in Fig.~\ref{SM_Fig4} (a) exhibits no distinct features, indicating that the system remains in the thermal equilibrium phase. Once $E_{\rm{MW}}$  exceeds the critical field $E_{\rm{c}}$, the spectrum progressively develops the hallmark signatures of the time crystal phase, including a fundamental frequency ($\sim$ 6 kHz) and several higher-order harmonics. The critical field here is approximately 1.64 mV/cm, which differs slightly from the value in Fig.~\ref{SM_Fig3} (b) due to a small variation in the optical detuning $\Delta$ between the two experimental sets. As $E_{\rm{MW}}$ increases further, both the fundamental and harmonic frequencies shift to higher values, accompanied by a significant increase in their spectral intensities, indicating enhanced robustness of the time crystal phase. Fig.~\ref{SM_Fig4} (b) displays the corresponding time crystal spectra at $E_{\rm{MW}} = 2.4$ (upper panel) and 3.5 mV/cm (lower panel), where the fundamental peak shifts from 8 kHz to 13 kHz, and the intensities of all peaks, including the harmonics, are markedly enhanced. 
    \begin{figure*}
        \centering
        \includegraphics[width=1\columnwidth]{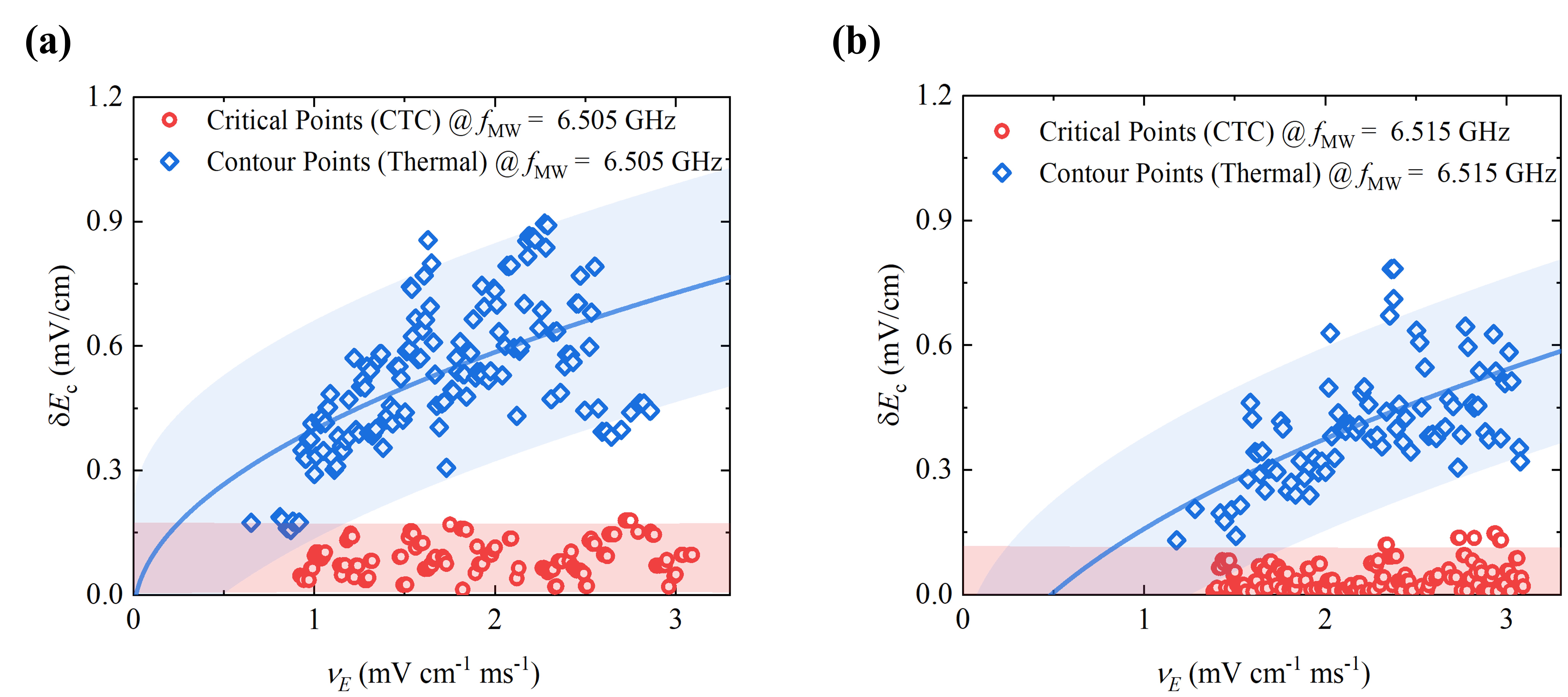}
        \caption{\textbf{Dependence of the critical electric field error $\delta E_c$ on the microwave electric field scanning rate $\nu_E$ at off-resonant MW frequencies.} (a) Experimental results for $f_{\text{MW}} = 6.505\ \text{GHz}$. The error $\delta E_c$ of the CTC phase remains nearly constant with a mean value of $0.09\ \text{mV}/\text{cm}$ (red band indicates the $95\%$ prediction interval), whereas for the thermal equilibrium phase, $\delta E_c$ increases and is fitted by $\delta E_c = a_{\text{Th}} \sqrt{\nu_E} + b_{\text{Th}}$ ($a_{\text{Th}} = 0.45$, $b_{\text{Th}} = -0.05$), with blue band showing the corresponding $95\%$ prediction interval. (b) Corresponding results at $f_{\text{MW}} = 6.515\ \text{GHz}$, showing a similar trend with an average $\delta E_c$ of $0.04~\text{mV}/\text{cm}$ for the CTC phase and a fit for the thermal phase yielding $a_{\text{Th}} = 0.52$, $b_{\text{Th}} = -0.36$.}
        \label{SM_Fig5}
    \end{figure*}
    In the main text, we give a comparison of the measurement error $\delta E_{\rm c}$ between the thermal equilibrium system and the CTC system under different scanning rates $\nu_E$ at resonance. It is observed that the measurement error of the CTC system remains consistently low and is almost unaffected by changes in $\nu_E$, whereas the error of the thermal system increases with higher scanning rates. This trend persists at other microwave frequencies, as confirmed by experimental results of $\delta E_{\rm c}$ at two additional $f_{\rm{MW}}$ values shown in Fig.~\ref{SM_Fig5}. Additionally, as $f_{\rm{MW}}$ shifts away from the resonance, the minimum $\nu_E$ required for the emergence of the CTC phase increases. Specifically, the value rises from 0.7 $\rm{mV~cm^{-1}~ms^{-1}}$ at $f_{\rm{MW}} = 6.5~\rm{GHz}$ in Fig. 5 (f) of the main text, to 0.9 $\rm{mV~cm^{-1}~ms^{-1}}$ at $f_{\rm{MW}} = 6.505~\rm{GHz}$ in Fig.~\ref{SM_Fig5} (a) here, and further to 1.4 $\rm{mV~cm^{-1}~ms^{-1}}$ at $f_{\rm{MW}} = 6.515~\rm{GHz}$ in Fig.~\ref{SM_Fig5} (b). This progression agrees with the results in Fig. 4 (c) of the main text.

\bibliography{SMref}